\documentclass[preprintnumbers,prd,amsmath,amssymb,superscriptaddress,nofootinbib]{revtex4}

\usepackage{graphicx}
\usepackage[colorlinks=true,citecolor=blue]{hyperref}

\newcommand{\myeqref}[1]{\eqref{#1}}

\newcommand{\rmi}{\mathrm{i}}
\newcommand{\rmd}{\mathrm{d}}

\newcommand{\rmL}{\mathrm{L}}
\newcommand{\rmH}{\mathrm{H}}
\newcommand{\rmcr}{\mathrm{cr}}
\newcommand{\rmmax}{\mathrm{max}}
\newcommand{\thetav}{\theta_\mathrm{v}}
\newcommand{\sfH}{\mathsf{H}}
\newcommand{\sfU}{\mathsf{U}}
\newcommand{\sfP}{\mathsf{P}}
\newcommand{\sfB}{\mathsf{B}}
\newcommand{\sfL}{\mathsf{L}}
\newcommand{\sfD}{\mathsf{D}}
\newcommand{\sfX}{\mathsf{X}}
\newcommand{\stH}{\tilde{\mathsf{H}}}
\newcommand{\stP}{\tilde{\mathsf{P}}}
\newcommand{\stD}{\tilde{\mathsf{D}}}
\newcommand{\diag}{\mathrm{diag}}
\newcommand{\Tr}{\mathrm{Tr}}
\newcommand{\dmsqr}{\Delta m^2}

\newcommand{\eff}{\mathrm{eff}}

\newcommand{\sync}{\mathrm{sync}}
\newcommand{\spl}{\mathrm{s}}
\newcommand{\GF}{G_\mathrm{F}}
\newcommand{\bfx}{\mathbf{x}}
\newcommand{\bfp}{\mathbf{p}}
\newcommand{\bhp}{\hat{\mathbf{p}}}
\newcommand{\vP}{\vec{P}}
\newcommand{\vH}{\vec{H}}
\newcommand{\vtH}{\vec{\tilde{H}}}
\newcommand{\vB}{\vec{B}}
\newcommand{\vL}{\vec{L}}

\newcommand{\vD}{\vec{D}}

\newcommand{\vfe}{\hat{e}^\mathrm{(I)}}
\newcommand{\vve}{\hat{e}^\mathrm{(V)}}
\newcommand{\tlh}{\tilde{h}}
\newcommand{\tlp}{\tilde{p}}
\newcommand{\bcdot}{\boldsymbol{\cdot}}
\newcommand{\bfK}{\mathbf{K}}
\newcommand{\Rcp}{R_\mathrm{coll}^+}
\newcommand{\Rcm}{R_\mathrm{coll}^-}

\begin{document}




\preprint{LA-UR 09-08309}
\preprint{INT-PUB 10-001}
\title{Collective Neutrino Oscillations}


\author{Huaiyu Duan}
\affiliation{Theoretical Division, Los Alamos National Laboratory, Los
  Alamos, NM 87545}
\author{George M.\ Fuller}
\affiliation{Department of Physics, University of California, San Diego, 
La Jolla, CA 92093}
\author{Yong-Zhong Qian}
\affiliation{School of Physics and Astronomy, University of Minnesota,
Minneapolis, MN 55455}


\begin{abstract}
We review the rich phenomena associated with
neutrino flavor transformation in the presence of neutrino self-coupling.
Our exposition centers on three collective neutrino oscillation scenarios:
a simple bipolar neutrino system that initially consists of
mono-energetic $\nu_e$ and $\bar\nu_e$; a homogeneous and
isotropic neutrino gas with multiple neutrino/antineutrino species and
continuous energy spectra; and a generic neutrino gas in an
anisotropic environment. We use each of these scenarios to illustrate
key facets of collective neutrino oscillations.
We discuss the implications of collective neutrino flavor oscillations
for core collapse supernova physics and for the prospects of
obtaining fundamental neutrino properties, e.g., the neutrino mass
hierarchy and $\theta_{13}$ from a future observed supernova neutrino
signal.
\end{abstract}

\maketitle

\section{Introduction}

\subsection{Neutrino mixing and astrophysics}

Neutrinos and the phenomena associated with neutrino flavor
transformation stand at the nexus of two exciting recent developments:
the success of experimental neutrino physics; and the tremendous
growth of astronomy and astrophysics. The former enterprise has given
key insights into neutrino mass and vacuum mixing and promises more
\cite{Camilleri:2008zz},
while the latter is providing fundamental cosmological parameters and
is revealing how structure and elemental abundances emerge and evolve
in the universe 
\cite{Timmes:1994nd,Abel:2001pr,Samtleben:2007zz,Steigman:2007xt}. 
Moreover, there is feedback between these subjects.
For example, observations of large scale structure and the cosmic
microwave background radiation currently provide our best limits on
the neutrino rest masses \cite{Hannestad:2006zg,Fuller:2008nt}. In
fact, both the 
early universe and the 
massive star core collapse and supernova explosion environments can be
dominated by neutrinos. Neutrino flavor transformation in each of
these environments may give insights into astrophysics and even
possibly into fundamental neutrino properties. Obtaining these
insights will require confident modeling of neutrino flavor evolution in
environments where neutrino-neutrino interactions produce vexing
nonlinearity. In any case, the neutrino mass and mixing data already
gathered from the experiments make a compelling case that we must solve
this problem.    

Experiments and observations to date have established that the
neutrino energy (mass) states $\vert \nu_i\rangle$ ($i=1,2,3$) are not
coincident with the weak interaction states $\vert
\nu_\alpha\rangle$ ($\alpha=e,\mu,\tau$). The relation between these
bases is given by $\vert\nu_\alpha\rangle = \sum_i U_{\alpha i}^*
\,\vert\nu_i\rangle$ \cite{Amsler:2008zzb}, where the
Maki-Nakagawa-Sakata (MNS) matrix 
elements $U_{\alpha i}$ are parameterized by three vacuum mixing
angles ($\theta_{1 2}$, $\theta_{2 3}$, $\theta_{1 3}$) and a
$CP$-violating phase $\delta$. Two of these are measured outright,
$\sin^2\theta_{ 1 2}\approx 0.31$ and $\sin^2\theta_{2 3} \approx
0.5$, while there is a firm upper limit on a third, $\sin^2\theta_{1
  3} < 0.04$ at $2\sigma$ \cite{Schwetz:2008er}.  
Observations of solar neutrinos show flavor conversion in the
$\nu_e\rightleftharpoons\nu_{\mu/\tau}$ channel with a characteristic
mass-squared splitting $\Delta m^2_\odot \approx 7.6\times
{10}^{-5}\,{\rm eV}^2$. Atmospheric neutrino measurements show near
maximal vacuum mixing in the $\nu_\mu\rightleftharpoons\nu_\tau$
channel with corresponding mass-squared splitting $\Delta m^2_{\rm
  atm} \approx 2.4\times {10}^{-3}\,{\rm eV}^2$. However, experiments
do not reveal the absolute neutrino rest masses $m_i$ ($i=1,2,3$), nor
do they show whether these neutrino mass eigenvalues are ordered in
the normal mass hierarchy ($m_3>m_2>m_1$) or in the inverted mass
hierarchy ($m_2>m_1>m_3$).  

Future terrestrial neutrino experiments 
\cite{Mena:2004sa,Guo:2007ug,Bandyopadhyay:2007kx,Ray:2008zz,Terri:2009zz,Collaboration:2009yc}
will be directed primarily towards
measuring $\theta_{1 3}$ and, if $\theta_{1 3}$ is big enough, the
neutrino mass hierarchy and possibly even the $CP$-violating phase
$\delta$. The planned reactor and long baseline experiments may be
able to measure $\theta_{1 3}$ if it satisfies $\sin^2\theta_{1 3} >
{10}^{-4}$. This limit is set ultimately by constraints on the
neutrino flux and detector mass. To find significantly larger neutrino
fluxes we must turn to cosmic sources, e.g., core collapse supernovae
\cite{Fuller:1998kb,Lunardini:2003eh,Beacom:2003nk,Cadonati:2000kq,Ahrens:2001tz,Sharp:2002as,Ikeda:2007sa,Autiero:2007zj}.

Stars more massive than $\sim 8\,M_\odot$ end their lives in
gravitational collapse and the production of a neutron star or, if
they are massive enough, a black hole remnant 
\cite{Colgate:1961aa,Woosley:2002zz,Woosley:2005yv,Janka:2006fh}.
Neutrinos play a role in nearly every aspect of
the evolution of these core collapse supernovae, from dominating
lepton number and 
entropy loss from the epoch of core carbon/oxygen burning onward,
to providing the bulk of energy and
lepton number transport during collapse itself, and even to providing
the heating necessary to engender convection 
\cite{Herant:1994dd,Burrows:1995ww,Buras:2005rp,Bruenn:2007bd,Ott:2008jb,Hammer:2009cn}
and, e.g., the
Standing Accretion Shock Instability (SASI) which may create an
explosion \cite{Blondin:2005wz,Blondin:2006yw}.

A key point is that gravitational collapse causes an appreciable
fraction of the rest mass of the Chandrasekhar-mass ($\sim
1.4\,M_\odot$) core to appear as seas of trapped neutrinos which 
subsequently diffuse out of the core on time scales of seconds
\cite{Bethe:1984ux,Bethe:1990ee,Arnett:1996aa}. At
core bounce, the energy in the neutrino seas trapped in the core is
$\sim {10}^{52}\,{\rm ergs}$, but by $\sim 10\,{\rm s}$ after core
bounce some ${10}^{53}\,{\rm ergs}$, or $\sim 10\%$ of the rest mass
of the core has been radiated away as neutrinos of all kinds. 
The emergent
neutrino and antineutrino energy spectra and fluxes at the neutrino
sphere vary with time post-core-bounce, but there will be
epochs where the energy spectra and/or the luminosities in the various
neutrino flavors will differ.  Moreover,
the charged-current, flavor specific neutrino interaction processes such as
$\nu_e+n\rightleftharpoons p+e^-$ and 
$\bar\nu_e+p\rightleftharpoons n+e^+$ are 
important both for energy and electron lepton number deposition as
well as for determining neutrino transport physics
\cite{Qian:1993dg,Horowitz:2001xf,Prakash:2001rx}.
It is therefore interesting and necessary to assess whether
interconversion of neutrino 
flavors in the supernova environment affects explosion physics and
neutrino-heated nucleosynthesis and how such flavor transformation
might affect a supernova neutrino burst signature in a terrestrial
detector.

\subsection{Brief history of collective neutrino oscillations
\label{sec:literature}}

Early studies of neutrino flavor transformation centered on solar neutrinos,
especially after it was recognized first by Wolfenstein
\cite{Wolfenstein:1977ue}, and then Mikheyev and Smirnov
\cite{Mikheyev:1985aa}, that the medium through which the neutrino 
propagates could alter the effective neutrino mass and mixing
properties.
Almost immediately after the Mikhyev-Smirnov-Wolfenstein (MSW)
mechanism was understood, it was
realized that the forward coherent scattering of neutrinos with other
neutrinos, somewhat misleadingly
called ``neutrino self-coupling'' or ``neutrino
self-interaction'', could generate a similar effect
\cite{Fuller:1987aa,Notzold:1988kx}. Subsequently, the effects of 
neutrino self-interaction were investigated independently in the core
collapse supernova and the early universe scenarios. The studies of
the supernova environment first focused on MSW-like effects
\cite{Fuller:1992aa,Fuller:1993ry,Qian:1993dg,Qian:1994wh,Qian:1995ua}.
However, it was pointed out that the neutrino self-interaction
potential is very different from the matter potential in that it can
have non-vanishing off-diagonal elements in the interaction basis
\cite{Pantaleone:1992xh,Pantaleone:1992eq}. 
Studies using the complete
neutrino self-interaction potential showed
that neutrinos could experience ``self-maintained coherent
oscillations'' or ``collective oscillations'' in
lepton-degenerate early universe scenarios
\cite{Samuel:1993uw,Kostelecky:1993yt,Kostelecky:1993dm,Kostelecky:1993ys,Kostelecky:1994dt,Samuel:1995ri,Kostelecky:1996bs,Pantaleone:1998xi,Pastor:2001iu,Dolgov:2002ab,Abazajian:2002qx}. 
By collective oscillations we mean a significant fraction of neutrinos
oscillate coherently with respect to each other.

\begin{figure*}
\begin{center}
$\begin{array}{@{}c@{\hspace{0.1in}}l@{}}
\includegraphics*[scale=0.5, keepaspectratio]{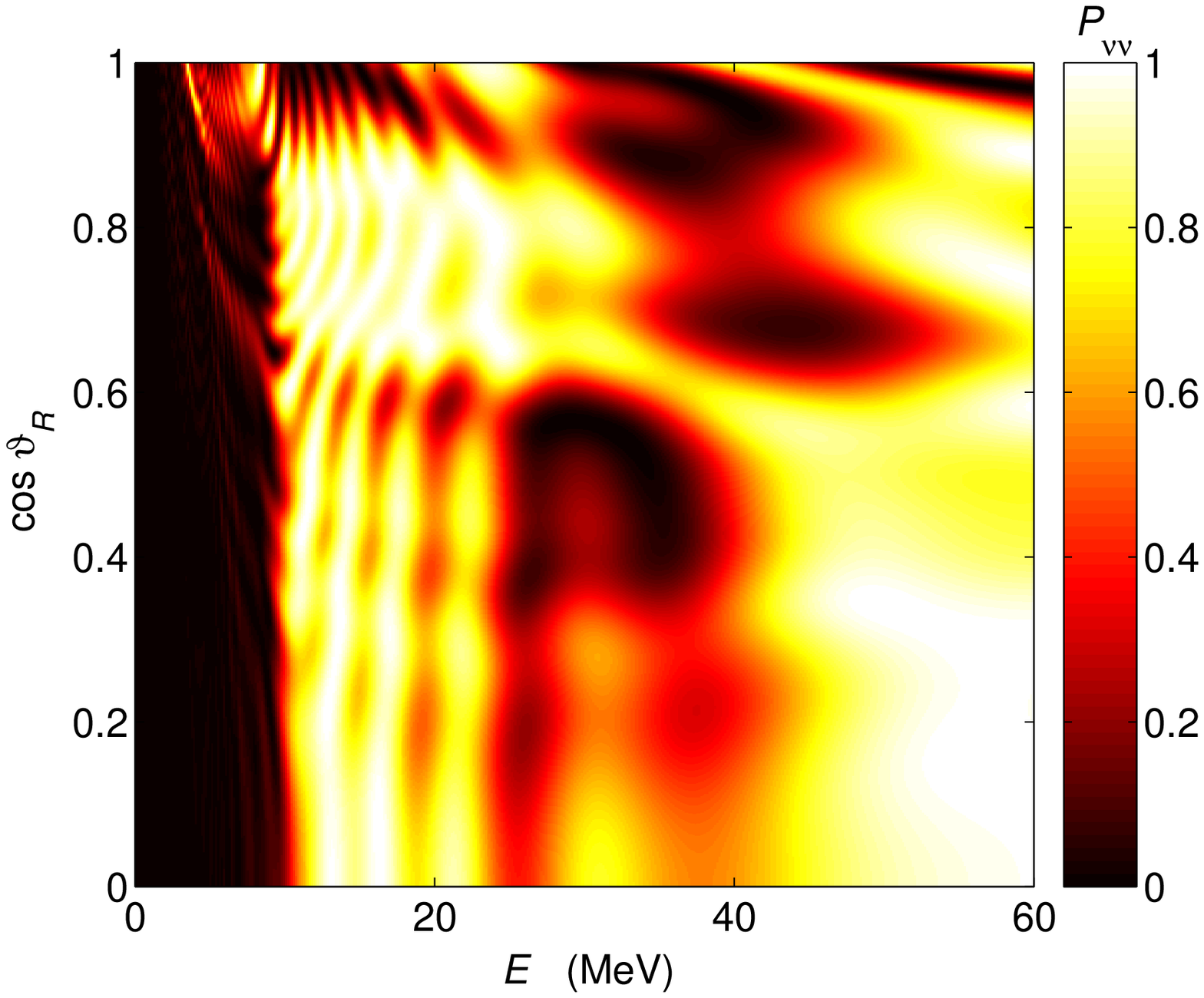} &
\includegraphics*[scale=0.5, keepaspectratio]{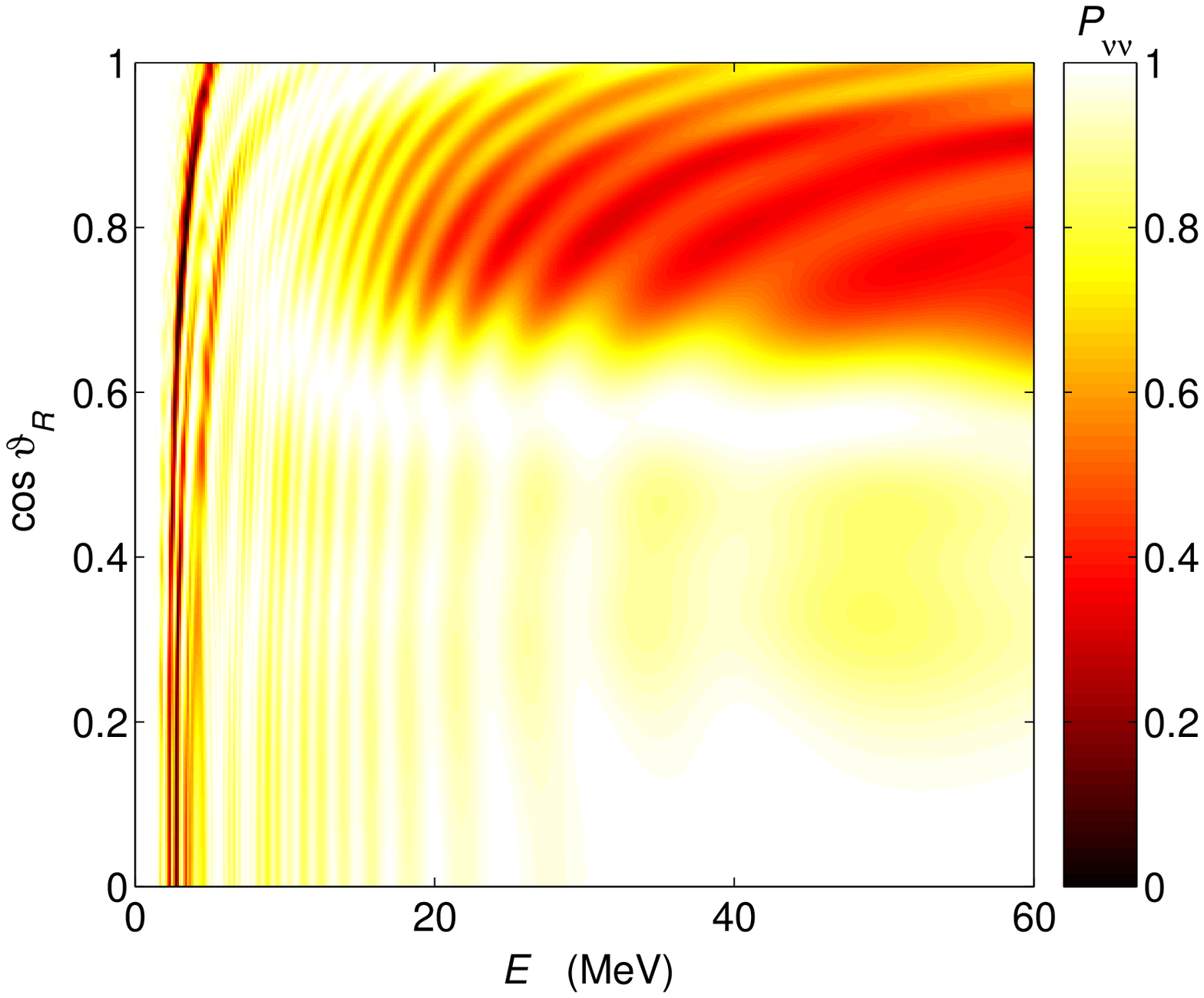} \\
\includegraphics*[scale=0.5, keepaspectratio]{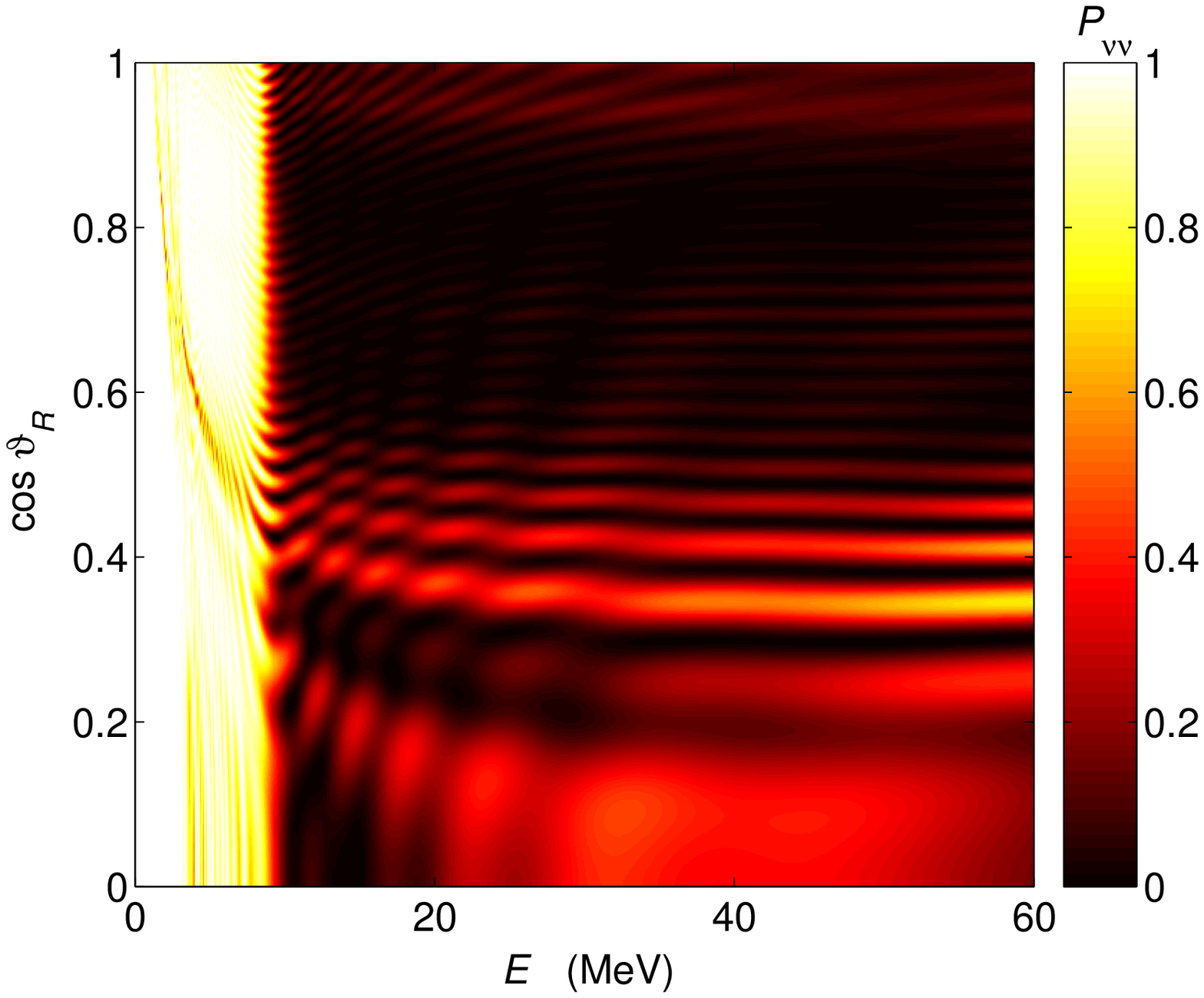} &
\includegraphics*[scale=0.5, keepaspectratio]{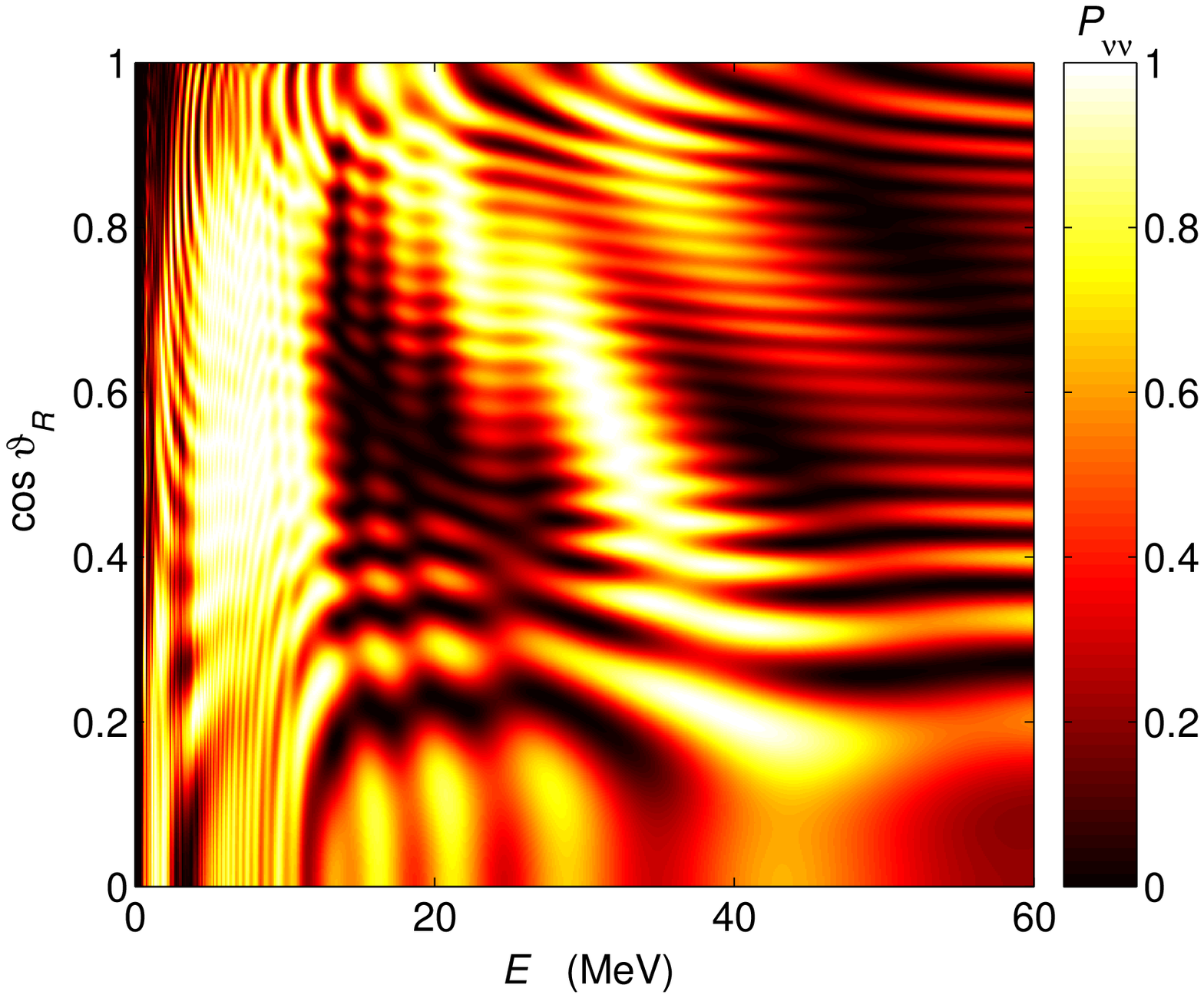}
\end{array}$
\end{center}
\caption{\label{fig:P-theta-E}
Survival probabilities $P_{\nu\nu}$ for neutrinos (left panels) and
antineutrinos (right panels) as functions of both neutrino energy $E$
and emission angle $\vartheta_R$  in a
numerical calculation using the neutrino bulb model (see
Figure~\ref{fig:nubulb}) and the two-flavor mixing scheme. 
The most prominent features on this figure are the approximately
angle-independent step-like changes in the neutrino survival
probabilities (left panels) --- these are the
 spectral swaps/splits discovered in Reference~\cite{Duan:2006an}.
The energy spectra of $\nu_e$ and $\nu_\mu$ with energy below (above)
$E\simeq9$ MeV in this calculation
are almost completely swapped in the upper (lower)
panels which employ the normal (inverted) neutrino mass hierarchy. 
Spectral swaps/splits are the result of a collective neutrino oscillation
mode, the precession mode, which is one of the main topics of this
review. The vertical fringes (the energy-dependent features in the
figure) are the result of MSW flavor transformation which is energy
dependent. The
horizontal fringes (the angle-dependent features) are the result of the
kinematic decoherence of bipolar neutrino oscillations
\cite{Raffelt:2007yz} (see Section~\ref{sec:decoherence}).
The movie version of these calculations is available in
Reference~\cite{Duan:2008eb}. 
Figure adapted from Figure~3 in
Reference~\cite{Duan:2006jv}.
Copyright 2006 by the American Physical Society.} 
\end{figure*}

Early research on MSW-like evolution, where the flavor off-diagonal
potentials were minimal, as well as the first paper
\cite{Pastor:2002we} to point out
that collective effects  could occur in supernovae
all focused on 
relatively large neutrino mass-squared differences. However, it was
eventually realized that neutrinos and antineutrinos could be
transformed collectively and simultaneously even for the small,
measured neutrino mass-squared differences
\cite{Balantekin:2004ug,Fuller:2005ae}. Additionally,
Reference~\cite{Duan:2005cp} showed that ordinary matter does not
necessarily suppress collective neutrino oscillations, at least in
homogeneous and isotropic environments.
It was then demonstrated that
collective neutrino oscillations indeed can occur in a spherically
symmetric supernova model \cite{Duan:2006jv,Duan:2006an}.
These works showed that the neutrino energy spectra would be modified
differently for the normal and inverted neutrino mass hierarchies (see
Figure~~\ref{fig:P-theta-E}). 
Over the past few years, many
papers have been written on the collective
neutrino oscillation phenomenon and its physical effects
\cite{Hannestad:2006nj,Raffelt:2007yz,Duan:2007mv,Raffelt:2007cb,EstebanPretel:2007ec,Duan:2007fw,Duan:2007bt,Fogli:2007bk,Raffelt:2007xt,Duan:2007sh,EstebanPretel:2007yq,Lunardini:2007vn,Dasgupta:2007ws,Duan:2008za,Dasgupta:2008cd,Dasgupta:2008my,Duan:2008eb,Sawyer:2008zs,Chakraborty:2008zp,Dasgupta:2008cu,EstebanPretel:2008ni,Gava:2008rp,Duan:2008fd,Raffelt:2008hr,Blennow:2008er,Fogli:2008fj,Lunardini:2008xd,Guo:2008mma,Minakata:2008nc,Liao:2009ic,Duan:2009cd,Liao:2009yz,Dasgupta:2009mg,Galais:2009wi,Fogli:2009rd,EstebanPretel:2009is,Gava:2009pj,Chakraborty:2009ej,Lazauskas:2009yh,Friedland:2010sc}.
Following this literature can be bewildering.
For example, the synchronized and bipolar neutrino
oscillations are frequently perceived as the two most important collective
neutrino oscillation modes in supernovae. However, numerical studies
suggest that supernova neutrinos are probably never synchronized
because of the non-vanishing matter density \cite{Duan:2007fw}, and
the bipolar oscillation is not a collective neutrino oscillation mode
in anisotropic environments such as supernovae \cite{Raffelt:2007yz}.

\subsection{Goal and organization of this review}

The purpose of this article is to provide a relatively short but
in-depth review of the properties of collective neutrino oscillations that are
reasonably well understood.
In particular, we wish to elucidate the physics behind the striking
features such as the swaps of supernova neutrino
energy spectra caused by collective oscillations as shown in
Figure~\ref{fig:P-theta-E}. 
Our strategy is to illustrate key facets of collective neutrino
oscillations by describing three,
increasingly more complex models. 

The rest of the review is organized as follows.
In Section~\ref{sec:mixing} 
we try to make the connection between
the ``wavefunction language'' and the ``spin language'', 
commonly used for studying non-collective and collective neutrino
oscillations, respectively. 
In Section~\ref{sec:bipolar}
we discuss the flavor evolution of a homogeneous and isotropic
neutrino gas that consists initially of mono-energetic pure $\nu_e$ and
$\bar\nu_e$. This simple model can be solved analytically and offers
important insights into the behavior of more complicated systems. In
Section~\ref{sec:isotropic-gas} we discuss homogeneous and isotropic
gases that consist of neutrinos with continuous energy spectra. We
explore 
adiabatic solutions to the neutrino flavor evolution equations and demonstrate
how the spectral swap/split phenomenon can be explained by one of these
solutions. In Section~\ref{sec:anisotropic-gas} we discuss some
of the important neutrino oscillation properties that are unique to
anisotropic environments and explain 
why the spectral swap/split phenomenon may occur in these environments
despite the anisotropy in neutrino fields. 
In Section~\ref{sec:supernovae} we apply the
current understanding of collective neutrino oscillation
phenomenon to supernova environments. In
Section~\ref{sec:conclusions} we give a summary and 
point out several issues in collective neutrino oscillations
that remain to be understood.

\section{Neutrino mixing in dense neutrino gases
\label{sec:mixing}}

\subsection{Equations of motion}
Here we focus on two-flavor neutrino mixing
scenarios, i.e., $\alpha=e,\mu$ and $i=1,2$, where $|\nu_\mu\rangle$
is a linear combination of the physical $|\nu_\mu\rangle$ and
$|\nu_\tau\rangle$.  
We do this for pedagogical purposes, although this is physically
justifiable because the physical $\nu_\mu$ and $\nu_\tau$ are nearly
maximally mixed in vacuum and experience nearly identical interactions
in the supernova environment \cite{Balantekin:1999dx,Caldwell:1999zk}.
We will discuss
collective neutrino oscillations with the full three-flavor mixing machinery
in Section~\ref{sec:three-flavor}. 
We consider only coherent neutrino flavor evolution, where the
effects of neutrino inelastic scattering and associated quantum
decoherence can be neglected. 
This will be generally applicable in the region well above the neutron
star in supernova models.
(Solution of the complete problem of neutrino flavor evolution with
both elastic and inelastic neutrino scattering would necessitate 
the use of the full quantum kinetic equations 
\cite{Prakash:2001rx,Sigl:1992fn,McKellar:1994aa,Strack:2005ux,Fuller:2008nt}.)
We also assume that neutrinos are relativistic
and that general
relativistic effects can be ignored.
With these assumptions, a mean-field Schr\"odinger-like equation
\begin{equation}
\rmi\frac{\rmd}{\rmd x}\psi
=\sfH\psi
\label{eq:eom-psi}
\end{equation}
is taken to describe flavor evolution along a neutrino world line 
 \cite{Halprin:1986pn}.
Here, we take $\hbar=c=1$,
 $x$ is the distance along the world line of the neutrino,
$\psi$ is the neutrino flavor wavefunction, taken to be a vector in 
(neutrino) flavor space, and 
\begin{equation}
\sfH= \frac{\dmsqr}{2E}\sfB + \lambda \sfL +\sfH_{\nu\nu}
= \frac{\dmsqr}{2E}\sfB + \sqrt{2}\GF n_e \sfL
+\sqrt{2}\GF
\int\!\rmd^3\bfp^\prime(1-\bhp\bcdot\bhp')
(\rho_{\bfp'} - \bar\rho_{\bfp'})
\label{eq:Hnunu}
\end{equation}
is the Hamiltonian. 
(See
References~\cite{Bell:2003mg,Friedland:2003eh,Sawyer:2005jk,Friedland:2006ke,Balantekin:2006tg}
for discussions of the 
applicability of the one-particle effective approximation assumed in
Equation~\myeqref{eq:eom-psi}.) 
The wavefunction for an antineutrino also obeys
Equation~\myeqref{eq:eom-psi}, but with the replacements 
$\lambda\rightarrow-\lambda$ and
$\sfH_{\nu\nu}\rightarrow -\sfH_{\nu\nu}^*$ in
Equation~\myeqref{eq:Hnunu}. 
The flavor space is spanned by the neutrino interaction basis or,
equivalently, spanned by the mass basis. We will use the word ``flavor'' in a
more general sense and we will avoid the phrase ``flavor state'',
which in the literature can be taken to 
mean either the interaction state $|\nu_\alpha\rangle$ 
or the flavor quantum state $|\psi\rangle$ . (We will use symbols
in the sans serif font, e.g.,
$\sfH$, to denote basis-dependent matrices in flavor space.)

The first term in Equation~\myeqref{eq:Hnunu} induces neutrino flavor
transformation because in the interaction basis
\begin{equation}
\sfB=\sfU\left(\frac{1}{2}\diag[-1,\,1]\right)\sfU^\dagger
= \frac{1}{2}\begin{bmatrix}
-\cos2\thetav & \sin2\thetav \\
\sin2\thetav & \cos2\thetav
\end{bmatrix}
\label{eq:B}
\end{equation}
is non-diagonal. In Equation~\myeqref{eq:B}, $\sfU$ is the MNS matrix,
and $\thetav$ is the so-called
vacuum mixing angle and is within the range $(0,\pi/4]$.
In Equation~\myeqref{eq:Hnunu} $\dmsqr$ is the 
mass-squared difference appropriate for $|\nu_2\rangle$ and $|\nu_1\rangle$, and
$E$ is the energy of the neutrino.
Here $\dmsqr>0$ and $\dmsqr<0$ correspond to
the normal neutrino mass hierarchy (NH)
and the inverted neutrino mass hierarchy (IH), respectively,
as discussed above for the full three-flavor mixing case.
Although this formalism is completely general, one of the most
interesting supernova cases is where $|\dmsqr|\simeq\dmsqr_\mathrm{atm}$
and $\thetav\simeq\theta_{13}\ll1$.

The second term in Equation~\myeqref{eq:Hnunu} arises from 
coherent neutrino-electron  forward exchange scattering
\cite{Wolfenstein:1977ue}.
In this term, which is referred to as the matter term,
 $\GF$ is the Fermi constant, $n_e$ is the net electron number
density, and $\sfL=\diag [1,0]$ in the interaction basis. 
Note that adding/subtracting a multiple of the identity matrix to/from $\sfH$
(e.g., the contribution of neutral-current neutrino-electron scattering)
gives only an overall phase to $\psi$ and, therefore, does not affect
neutrino oscillations. 

The last term in Equation~\myeqref{eq:Hnunu} stems from 
coherent neutral-current neutrino-neutrino forward exchange scattering
\cite{Fuller:1987aa,Notzold:1988kx,Fuller:1992aa,Pantaleone:1992xh}, where 
$\bhp$ and $\bhp'$ are the unit vectors for the propagation
directions of the test neutrino and the background neutrino 
or antineutrino, respectively. In the interaction basis,
at location $\bfx$ and at time $t$,
the (flavor) density matrices for neutrinos
and antineutrinos with momentum $\bfp'$ 
and with our assumptions 
can be written as
\begin{subequations}
\begin{align}
[\rho_{\bfp'}(t,\bfx)]_{\alpha\beta} 
&= \sum_{\nu'} n_{\nu',\bfp'} (t,\bfx)
\langle\nu_\alpha|\psi_{\nu',\bfp'}(t,\bfx)\rangle
\langle\psi_{\nu',\bfp'}(t,\bfx)|\nu_\beta\rangle, \\
[\bar\rho_{\bfp'}(t,\bfx)]_{\beta\alpha} 
&= \sum_{\bar\nu'} n_{\bar\nu',\bfp'} (t,\bfx)
\langle\bar\nu_\alpha|\psi_{\bar\nu',\bfp'}(t,\bfx)\rangle
\langle\psi_{\bar\nu',\bfp'}(t,\bfx)|\bar\nu_\beta\rangle,
\label{eq:rhobar}
\end{align}
\end{subequations}
respectively,
where $|\psi_{\nu'(\bar\nu'),\bfp'}\rangle$ is the state of a neutrino
$\nu'$ (antineutrino $\bar\nu'$) with momentum $\bfp'$, and 
$n_{\nu',\bfp'}$ ($n_{\bar\nu',\bfp'}$) is the corresponding number
density of the neutrino (antineutrino). 
Note that the order of the indices on the matrix representation 
of $\bar\rho$ in Equation~\myeqref{eq:rhobar} 
follows the convention in
Reference~\cite{Sigl:1992fn}. The advantage of this definition is that
$\bar\rho$ will transform in the same way 
as does $\rho$  when transforming from the interaction basis to the mass
basis or vice versa.

\subsection{Neutrino flavor polarization vector}
It becomes more difficult to
analyze neutrino oscillations using the wavefunction
formalism when $\sfH_{\nu\nu}$ is significant.
 This is because
$\sfH_{\nu\nu}$ is a sum of density matrices
which involve bilinear forms of the wavefunctions. 
The density matrices
contain all the physical information for the neutrino mixing problem. 
The diagonal elements of the density matrices give the number densities
of neutrinos in the weak interaction states or the
mass states, 
depending on the basis used, and the off-diagonal elements of the
density matrices contain the neutrino mixing information.
For simplicity let us first take a working example,
 a homogeneous and isotropic neutrino
gas whose flavor content can vary with time $t$. 
Homogeneity and isotropy implies that the factor $(1-\bhp\bcdot\bhp')$ in
Equation~\myeqref{eq:Hnunu} averages to $1$. Therefore, the neutrino
propagation direction does not matter, and
$\rho_\bfp(t,\bfx)\rightarrow\rho_E(t)$ and
$\bar\rho_\bfp(t,\bfx)\rightarrow\bar\rho_E(t)$.
These density matrices obey the equations of
motion (EoM):
\begin{subequations}
\begin{align}
\rmi\dot\rho_E &= 
\left[\frac{\dmsqr}{2E}\sfB + \lambda \sfL
+\sqrt{2}\GF \int_0^\infty\rmd E'(\rho_{E'} - \bar\rho_{E'}),\,
\rho_E\right],\\
\rmi\dot{\bar\rho}_E &= 
\left[-\frac{\dmsqr}{2E}\sfB + \lambda \sfL
+\sqrt{2}\GF \int_0^\infty\rmd E'(\rho_{E'} - \bar\rho_{E'}),\,
\bar\rho_E\right].
\end{align}
\label{eq:eom-rho}
\end{subequations}

Because $\rho_E$ and $\bar\rho_E$ are $2\times2$ Hermitian matrices,
they can be mapped into vectors in a three-dimensional Euclidean space
which we will also call flavor space. We
define the components of the (neutrino flavor) polarization vector 
$\vec{P}_\omega$ to be
\begin{equation}
P_{\omega,a}=\left(\frac{1}{n_{\nu}}\right)
\left(\frac{|\dmsqr|}{2\omega^2}\right)\times
\left\{\begin{array}{ll}
\Tr(\rho_E\, \sigma_a) & \text{ for neutrino},\\
-\Tr(\bar\rho_E\, \sigma_a) &\text{ for antineutrino},
\end{array}\right.
\label{eq:vP}
\end{equation}
where the (angular) vacuum oscillation frequency is
$\omega=\frac{\dmsqr}{2E}$ for neutrinos and 
$\frac{\dmsqr}{(-2E)}$ for antineutrinos,
and $\sigma_a$ ($a=1,2,3$) are the Pauli matrices.
(We use symbols with the vector hat, e.g., $\vP$, to denote a vector
in flavor space and symbols in the bold font, e.g., $\bfp$, to denote
a vector in physical three-dimensional coordinate space.)
We note that $\vP_\omega$ can be normalized by an arbitrary factor.
For example, if $\vP_\omega$ is normalized to unity, then
$\rho_E\propto(1+\frac{1}{2}\vP_\omega\cdot\vec{\sigma})$ for neutrinos
and $\bar\rho_E\propto(1-\frac{1}{2}\vP_\omega\cdot\vec{\sigma})$ for
antineutrinos. In Equation~\myeqref{eq:vP} we
defined $\vP_\omega$ to be normalized by $n_{\nu}$, the initial total
number density of a 
certain neutrino species $\nu$ which is chosen to be $\bar\nu_e$ in the
rest of the review. However, we usually take $\mu=\sqrt{2}\GF n_\nu$
as a measure of the strength of neutrino
self-interaction. Therefore, in some cases it is more
appropriate to normalize $\vP_\omega$ by the number density of
other neutrino species, e.g., when $n_{\bar\nu_e}$ is negligible.  
If neutrinos and antineutrinos are
all in the interaction states, then 
$\vP_\omega\propto (n_{\nu_e,\omega}-n_{\nu_\mu,\omega})\vfe_3$ for neutrinos and 
$\vP_\omega\propto
-(n_{\bar\nu_e,\omega}-n_{\bar\nu_\mu,\omega})\vfe_3$ for
antineutrinos, where $n_{\nu,\omega}$
($\nu=\nu_e,\nu_\mu,\bar\nu_e,\bar\nu_\mu$) is the corresponding
number density in the neutrino or antineutrino mode $\omega$,
and $\vfe_3$ is one of the interaction basis vectors.

\begin{figure}
\begin{center}
$\begin{array}{@{}c@{\hspace{0.6 in}}c@{}}
\includegraphics*[scale=0.6, keepaspectratio]{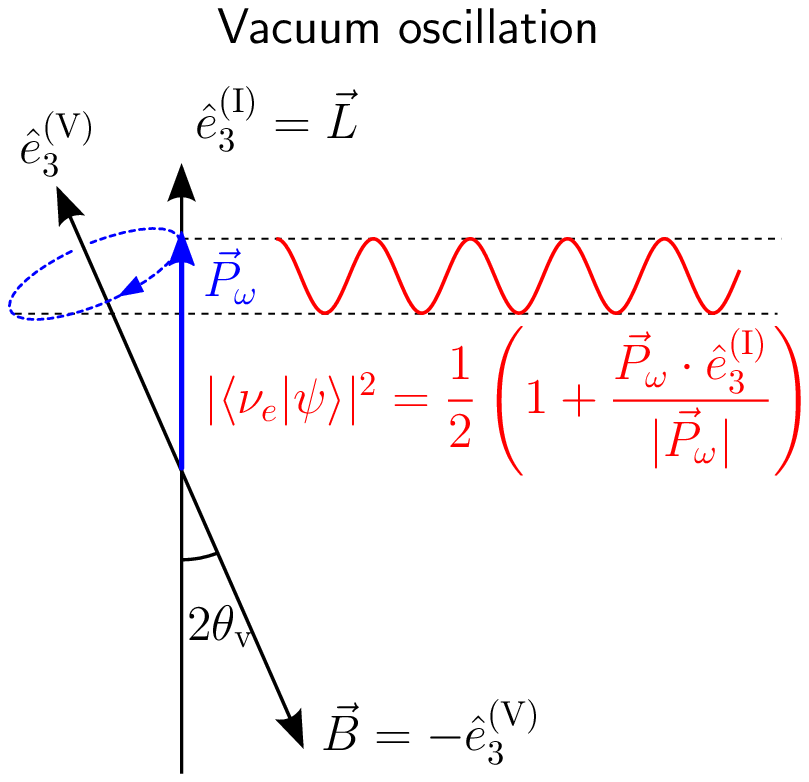} &
\includegraphics*[scale=0.6, keepaspectratio]{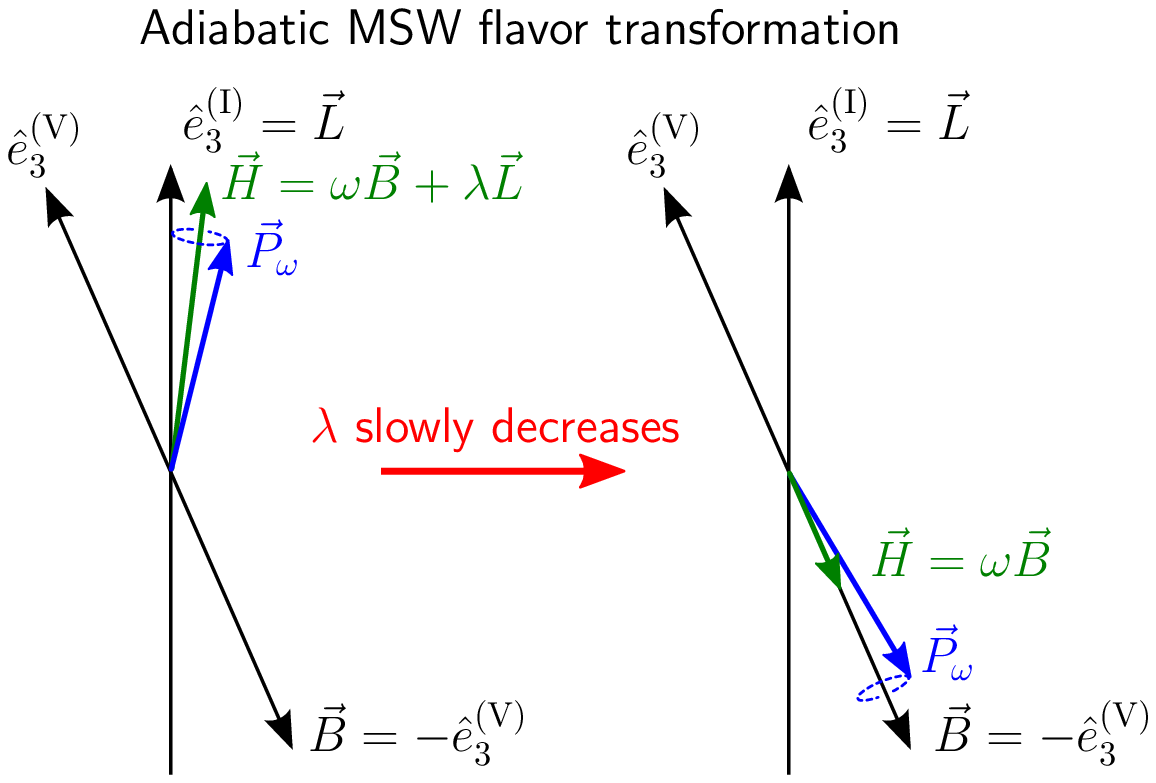} 
\end{array}$
\end{center}
\caption{\label{fig:vacuum-MSW}%
Geometric pictures for vacuum oscillations and for MSW flavor
transformation. The flavor Hilbert space spanned by
$|\nu_\alpha\rangle$ ($\alpha=e,\mu$) or
$|\nu_i\rangle$ ($i=1,2$) can be mapped onto the flavor Euclidean
space that is spanned by  $\vfe_a$ or $\vve_a$
($a=1,2,3$). The interaction basis vectors
$\vfe_{1,3}$ can be obtained by rotating
 the (vacuum) mass basis vectors $\vve_{1,3}$ 
 by $2\thetav$ about $\vfe_2=\vve_2$ . 
The left panel shows the polarization vector $\vP_\omega$,
which describes a neutrino initially in $|\nu_e\rangle$,
and its precession about $\vB$. The
projection of this precession motion onto the $\vfe_3$ axis represents
the flavor oscillation of the neutrino. The right panel shows the
precession of $\vP_\omega$ in the presence of ordinary matter. If the
matter density varies slowly, the angle between $\vP_\omega$ and
$\vH=\omega\vB+\lambda\vL$ remains constant. This represents adiabatic
MSW flavor transformation of the neutrino. See also
Reference~\cite{Kim:1987bv} for a 
more detailed discussion on this geometric interpretation.}
\end{figure}

Since the Pauli matrices are traceless,  the
trace of the density matrix is not contained in the
polarization vector. According to Equation~\myeqref{eq:eom-rho} 
the traces of the density matrices do not
change with time. (This corresponds to one of our assumptions that
neutrinos are not created or annihilated.)
These terms can be easily reintroduced, e.g., for the calculation of
 the neutrino energy spectra.
Using Equations~\myeqref{eq:eom-rho} and \myeqref{eq:vP} it can be shown that
\begin{equation}
\dot{\vP}_\omega
=(\omega\vB + \lambda\vL + \mu\vD)\times\vP_\omega,
\label{eq:eom-vP}
\end{equation}
where vectors $\vB=\Tr(\sfB\,\vec{\sigma})$ and
$\vL=\Tr(\sfL\,\vec{\sigma})$ are parallel to the (vacuum) mass and
interaction basis vectors $\vve_3$ and $\vfe_3$, respectively
(see Figure~\ref{fig:vacuum-MSW}), and $\vD =
\int_{-\infty}^\infty\vP_\omega\rmd\omega$ 
is the total polarization vector. 
In the absence of neutrino self-coupling, $\vP_\omega$ can be thought
of as a ``magnetic spin''. In this analogy the ``magnetic spin'' is
coupled to two ``magnetic fields'', 
$\vB$ and $\vL$, with gyromagnetic ratios $-\omega$ and $-\lambda$,
respectively. (Note that a real magnetic spin
$\mathbf{s}$ with the gyromagnetic ratio $\gamma$ in the presence of
magnetic field $\mathbf{B}$ obeys EoM
$\dot{\mathbf{s}}=-\gamma\mathbf{B}\times\mathbf{s}$.) Equivalently,
$\vP_\omega$ behaves like a ``magnetic spin'' coupled to the total
``magnetic field'' $\vH=\omega\vB+\lambda\vL$ with ``gyromagnetic
ratio'' $-1$. This picture allows geometric interpretations for both
vacuum oscillations and MSW
flavor transformation 
(see Figure~\ref{fig:vacuum-MSW}).

A few comments about the polarization vector notation are in order
before we consider neutrino self-coupling.
(a) The polarization vector notation is fully equivalent to the
neutrino flavor isospin (NFIS) notation \cite{Duan:2005cp} where the
NFIS for a neutrino 
or antineutrino is $\vec{s}_\omega=\frac{1}{2}\frac{\vP_\omega}{|\vP_\omega|}$.
(b) In Equation~\myeqref{eq:vP} $\vP_\omega$ is defined with a
minus sign for the antineutrino. Although the physics is not changed by
the choice of notation, our definition of $\vP_\omega$ is convenient in
analyzing collective neutrino  
oscillations when both neutrinos and antineutrinos are present (see
Section~\ref{sec:synchronization}). This notation has been adopted in 
most recent literature.

\subsection{Synchronized neutrino oscillations, 
corotating frames and matter effects
\label{sec:synchronization}} 

The neutrino self-coupling is represented as the coupling between
polarization vectors in
Equation~\myeqref{eq:eom-vP}. To have a feeling for the effect of this
coupling, let us 
consider a homogeneous and isotropic neutrino gas with $\lambda=0$ and
$\mu=\mathrm{const}$. Using Equation~\myeqref{eq:eom-vP} it can be shown
that the ``total energy of the magnetic spins'' 
\begin{equation}
\mathcal{E}=\int_{-\infty}^\infty\omega(\vP_\omega\cdot\vB)\rmd\omega
+\frac{\mu}{2}\vD^2
\label{eq:energy}
\end{equation}
is constant in this case \cite{Duan:2005cp},
where the first term is the ``total energy'' of coupling between
the ``magnetic field'' and ``spins'', and the second term is the
``total spin-spin coupling energy''.
If $\mu$ is large, then
$|\vD|\simeq\sqrt{2\mathcal{E}/\mu}$ is approximately constant. This
implies that a 
dense neutrino gas can experience ``self-maintained coherent
oscillations'' \cite{Kostelecky:1994dt}.  For example, if
a dense neutrino gas  consists initially of neutrinos of the same
flavor (so that all $\vP_\omega$ are initially aligned), then the
flavor evolution of these neutrinos is coherent
(i.e., all $\vP_\omega$ remain aligned)
 even when they have different energies. This phenomenon
has been termed synchronized neutrino oscillations, because all 
neutrinos (and antineutrinos) in such a system oscillate
collectively with (angular) frequency $\Omega_\sync$. 
The synchronized oscillation frequency is an average of all
$\omega$'s \cite{Pastor:2001iu}:
\begin{equation}
\Omega_\sync=
|\vD|^{-2}
\int_{-\infty}^\infty(\vD\cdot\vP_\omega)\omega\rmd\omega.
\label{eq:Omega-sync}
\end{equation}

Consider another neutrino system that is similar to the synchronized
system discussed above except that the oscillation frequency
$\omega$ of each neutrino or antineutrino is shifted by a common value
$\omega_0$. Any polarization vector,
say $\vP_\omega$, in this system should move in a way similar to
$\vP_{\omega-\omega_0}$ in the synchronized system, except for an
extra precession about $\vB$ with frequency $\omega_0$. In
other words, this system behaves just like the synchronized
system in a reference frame that rotates
about $\vB$ with  
frequency $\omega_0$. Indeed, in this non-inertial corotating frame
\cite{Duan:2005cp}, each
polarization vector is coupled to a non-physical field
$-\omega_0\vB$ and Equation~\myeqref{eq:eom-vP} (with $\lambda=0$)
becomes
\begin{equation}
\dot{\vP}_\omega = [(\omega-\omega_0)\vB+\mu\vD]\times\vP_\omega.
\end{equation}
Therefore, this neutrino system experiences synchronized
flavor transformation, just like the one discussed above, but with
$\Omega_\sync$ shifted by $\omega_0$. 
Because the oscillation frequency $\omega$ can be shifted to any
value by using an appropriate corotating frame, the criterion for
synchronization 
should not be $\mu\gg|\langle\omega\rangle|$ but
rather $\mu\gg\Delta\omega$ \cite{Duan:2005cp}, where
$\langle\omega\rangle$ and $\Delta\omega$ are the average value and the
spread in $\omega$ for the neutrino system, respectively.
Note that the reversal of the direction of $\vP_\omega$ for the
antineutrino (Equation~\myeqref{eq:vP}) is important when the corotating
frame is used. 
If this reversal is not incorporated into the definition of
$\vP_\omega$, then the direction of $\vP_\omega$ also changes if the sign of the
corresponding vacuum 
oscillation frequency changes on transformation to the corotating frame.

Note that if $\vB$ is 
parallel to $\vL$, the matter effect can be
completely removed by transforming to an appropriate corotating frame.
For a general case, on transforming to the corotating frame
Equation~\myeqref{eq:eom-vP} becomes
\begin{equation}
\dot{\vP}_\omega = (\omega\vB+\mu\vD)\times\vP_\omega
\quad\text{and}\quad
\dot{\vB} = -\lambda \vL\times\vB.
\label{eq:vP-eom-nomatt}
\end{equation}
The matter effect does not disappear here, but rather causes $\vB$ to
rotate in the corotating frame. 
If the matter density is
large ($\lambda\gg|\omega|$), however, the fast-rotating $\vB$ in
Equation~\myeqref{eq:vP-eom-nomatt} can be replaced by
$(\vB\cdot\vL)\vL$ for 
collective neutrino oscillations \cite{Duan:2005cp}. In other words,
for collective neutrino oscillations and in the presence of large matter
density, matter effects may be ``ignored''
and the effective neutrino mixing parameters become $\theta_\eff\simeq0$
and $\dmsqr_\eff=\dmsqr\cos2\thetav$.

\subsection{Solving for supernova neutrino flavor evolution}

\begin{figure}
\begin{center}
\includegraphics*[width=0.45\textwidth, keepaspectratio]{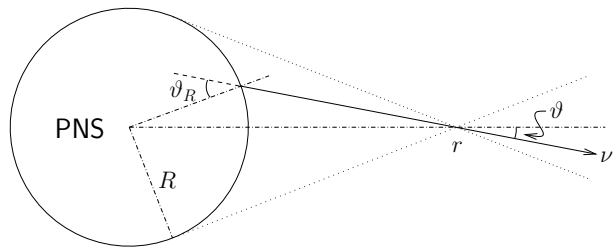}
\end{center}
\caption{\label{fig:nubulb}%
The geometric layout of the neutrino bulb
model. In this model all neutrinos are emitted half-isotropically from the 
surface (neutrino sphere) of the PNS which has radius $R$. 
Spherical symmetry and isotropic emission on the neutrino sphere 
imply that all neutrinos with the same initial flavor, energy and
emission angle $\vartheta_R$ have identical flavor evolution histories.
In this model the neutrino polarization vector
$\vP_{\omega,\vartheta}(r)$ is uniquely determined by $\omega$,
$\vartheta$ (or $\vartheta_R$) and $r$.
Here $\vartheta$ is the angle at radius $r$ between the neutrino
trajectory direction and the radial direction.
Figure adapted from Figure~1 in Reference~\cite{Duan:2006an}.
Copyright 2006 by the American Physical Society.}
\end{figure}

Although collective neutrino oscillations may occur in any
environment where neutrino fluxes are significant, recent studies of this
phenomenon have focused on the core collapse supernova
environment. 
The supernova environment is far more complex than the early universe,
in part because of its inhomogeneity and anisotropy.
This complexity is enhanced because
neutrinos of
different flavors and energies and propagating in different directions
are coupled by neutrino self-interaction \cite{Qian:1994wh}. Full
simulations of neutrino 
oscillations with neutrino self-interaction in a general supernova
environment are beyond current numerical capabilities.
Here we briefly discuss two schemes commonly
used in investigating collective neutrino oscillations.
Both of these schemes employ the neutrino bulb model where the supernova
environment is spherically symmetric 
around the center of the proto-neutron star (PNS)
(see Figure~\ref{fig:nubulb}). 
The polarization vectors in this model obey EoM
\begin{equation}
\cos\vartheta\frac{\rmd}{\rmd r} \vP_{\omega,\vartheta}(r)
= [\omega\vB + \lambda(r)\vL + \vH_{\nu\nu,\vartheta}(r)]\times
 \vP_{\omega,\vartheta}(r),
\label{eq:eom-bulb}
\end{equation}
where 
\begin{equation}
\vH_{\nu\nu,\vartheta}(r)=\sqrt{2}\GF n_{\bar\nu_e}(R)
\int_{-\infty}^\infty\rmd\omega'
\int_{\cos\vartheta_\mathrm{max}}^1\rmd(\cos\vartheta')\,
(1-\cos\vartheta\cos\vartheta')\vP_{\omega',\vartheta'}(r).
\label{eq:Hnunu-bulb}
\end{equation}
Note that the maximum value of $\vartheta(r)$ is
$\vartheta_\rmmax=\arcsin(R/r)$ in the neutrino bulb model (see
Figure~\ref{fig:nubulb}). 
Also note that here we choose to normalize $\vP_\omega$ by
$n_{\bar\nu_e}(R)=L_{\bar\nu_e}/(2\pi R^2\langle
  E_{\bar\nu_e}\rangle)$, the total
number density of $\bar\nu_e$ at the neutrino sphere,
where $L_{\bar\nu_e}$ and $\langle E_{\bar\nu_e}\rangle$ are the
energy luminosity and the spectrum-averaged energy of $\bar\nu_e$ at
the neutrino sphere, respectively. 
For the neutronization-burst epoch where the $\nu_e$ flux is much larger
than the fluxes of all other neutrino species,
$\vP_\omega$ should be normalized by $n_{\nu_e}(R)$ instead.
In typical numerical simulations,
 $R$ is in the range of $10$--$60$ km,
and the luminosities and the average energies of neutrinos are in the
ranges of $10^{50}$--$10^{53}$ erg/s and $10$--$30$ MeV,
respectively. 

Equation~\myeqref{eq:eom-bulb} can be solved numerically without any
further assumptions. This is the ``multi-angle scheme''.
The other scheme is the so-called ``single-angle scheme''. In this latter
scheme it is assumed that $\vP_{\omega,\vartheta}(r)=\vP_{\omega}(r)$
is the same for different neutrino trajectories. There are
several variants of the 
single-angle scheme. These variants lead to qualitatively similar
results. In one 
of the variants $\vP_\omega(r)$ is computed along the radial
direction ($\vartheta_R=0$) and Equation~\myeqref{eq:eom-bulb}
becomes
\begin{equation}
\frac{\rmd }{\rmd r}\vP_{\omega}(r)
= [\omega\vB + \lambda(r)\vL + 
\sqrt{2}\GF n_{\bar\nu_e} \mathcal{D}(r/R) \vD] \times
 \vP_{\omega}(r),
\label{eq:eom-single}
\end{equation}
where the geometric factor 
$\mathcal{D}(r/R)=\frac{1}{2}\left[1-\sqrt{1-(\frac{R}{r})^2}\right]^2$
partially accounts for the angle effect and geometric dilution of the
neutrino fluxes in the neutrino bulb model. Comparing
Equations~\myeqref{eq:eom-vP} and \myeqref{eq:eom-single}, it is clear
that the flavor evolution of 
neutrinos in the single-angle scheme is equivalent to that of
 a homogeneous and
isotropic neutrino gas expanding with ``time'' $r$.
In this analogy the strength of the neutrino 
self-coupling is $\mu(r)=\sqrt{2}\GF n_{\bar\nu_e}(R) \mathcal{D}(r/R)$.
The radial direction is a rather special direction in the neutrino
bulb model. In another variant of the single-angle scheme it is assumed
that all neutrinos are emitted with $\vartheta_R=\pi/4$
\cite{EstebanPretel:2007ec}. Alternatively, $\vH_{\nu\nu,\vartheta}$ can be
averaged over neutrino trajectories 
\cite{Duan:2006an,Dasgupta:2008cu}. Each of these variants also leads to
Equation~\myeqref{eq:eom-single} when $r\gg R$. However, in this limit
$\mathcal{D}(r/R)\rightarrow\frac{1}{4}\left(\frac{R}{r}\right)^4$
for these variants instead of $\frac{1}{8}(\frac{R}{r})^4$ in the
single-angle scheme which employs the radial trajectory. These variants
 can improve the agreement between the results of the
single-angle and multi-angle calculations \cite{Dasgupta:2008cu}. 

Although the single-angle scheme is frequently used for its simplicity,
it must be emphasized that it misses some of the important properties
(e.g., anisotropy) of the
neutrino bulb model and can lead to incorrect results when such properties
play important roles in collective neutrino oscillations (see Section
\ref{sec:anisotropic-gas}).

\section{Simple bipolar neutrino systems\label{sec:bipolar}}

\subsection{Bipolar systems and the flavor pendulum}

To illustrate another important example of collective neutrino
oscillations, let us consider a homogeneous and isotropic gas that
initially consists of mono-energetic $\nu_e$ and $\bar\nu_e$. 
This neutrino system is represented by two polarization
vectors, $\vP_\omega$ and $\vP_{-\omega}$ for the neutrino and the
antineutrino, respectively. We assume that 
$\lambda=0$, $\mu$ is fixed, and that
$|\vP_\omega|=(1+\varepsilon)|\vP_{-\omega}|$, where $\varepsilon$ is
the fractional excess of neutrinos over antineutrinos. At $t=0$,
$\vP_\omega$ points in the direction of $\vfe_3$, which is tilted away
from $\vve_3$ by $2\thetav$, and $\vP_{-\omega}$ points in the
direction opposite to $\vP_\omega$.
Neutrino systems that are represented by two nearly oppositely
directed polarization vector groups are called ``bipolar systems''.
Using the corotating frame technique,
the discussion in this section easily can be  applied to,
e.g., a gas consisting  initially 
of $\nu_e$ and $\nu_\mu$ with energies $E_{\nu_e}\neq E_{\nu_\mu}$.

A peculiar case is where $\varepsilon=0$ and $\vP_\omega$ initially is 
aligned with $\vve_3=-\vB$ (i.e., $\thetav=0$). Using energy
conservation 
(Equation~\myeqref{eq:energy}) it can be shown \cite{Duan:2005cp} that
if $\omega>0$, neither 
$\vP_\omega$ nor $\vP_{-\omega}$ will move, and so the initial
configuration of the system is absolutely stable. On the other hand,
if $\omega<0$ and $\mu\gg|\omega|$, $\vP_\omega$ and $\vP_{-\omega}$
can nearly swap their directions (but with a slight bend towards each
other). This implies that the initial configuration of the system is
unstable. Therefore,
the $\nu_e$--$\bar\nu_e$ system may experience insignificant
flavor oscillations when $\dmsqr>0$ and $\thetav\ll1$. 
However, this system can experience significant flavor oscillations
when $\dmsqr<0$ and $\thetav\ll1$. 
Collective neutrino oscillations of this kind are known as ``bipolar
oscillations''. 

The EoM of the simple bipolar system
\begin{equation}
\dot{\vP}_\omega = (\omega\vB+\mu\vD)\times\vP_\omega,
\quad\text{and}\quad
\dot{\vP}_{-\omega} = (-\omega\vB+\mu\vD)\times
\vP_{-\omega}
\label{eq:eom-bipolar}
\end{equation}
have been solved analytically
\cite{Kostelecky:1994dt,Samuel:1995ri}. Instead of presenting this
solution, let us rewrite \cite{Hannestad:2006nj} 
Equation~\myeqref{eq:eom-bipolar} as
\begin{equation}
\dot{\vD} = \mu^{-1}\vec{q}\times\vec{g}
\quad\text{and}\quad
\vD =\mu^{-1}\vec{q}\times\dot{\vec{q}} + \sigma_\mathrm{s}\vec{q},
\label{eq:eom-pendulum}
\end{equation}
where
$\vec{q}=\vec{Q}/|\vec{Q}|=(\vP_\omega-\vP_{-\omega}-\frac{\omega}{\mu}\vB)/|\vec{Q}|$,
$\vec{g}=-\mu\omega |\vec{Q}|\vB$, and
$\sigma_\mathrm{s}=\vec{q}\cdot\vD$ is constant.
Equation~\myeqref{eq:eom-pendulum}
 describes the motion of a fictitious gyroscopic pendulum, or ``flavor
 pendulum'',  with total
angular momentum $\vD$ in a uniform gravitational field where the acceleration
of gravity is $\vec{g}$. The pendulum consists of a massless rod with
a point particle of mass $\mu^{-1}$ and spin $\sigma_\mathrm{s}$
attached to the end of the rod at position $\vec{q}$. 
We note that the flavor pendulum that represents a symmetric bipolar system
($\varepsilon=0$) has no internal spin. The stable and unstable
configurations of the system discussed above correspond to the
lowest and highest positions, respectively, that the pendulum can reach.

Generally, the flavor pendulum can experience two kinds of motion:
a precession about the $\vB$ axis and a nutation around the
average precession track. The 
nutation motion corresponds to bipolar neutrino
oscillations. However, like a child's top, the flavor
pendulum can ``defy gravity'' and precess almost uniformly if
$\mu$ is large enough. In particular, the flavor pendulum can become a
``sleeping top'' and will not fall from its
highest position if \cite{Hannestad:2006nj,Duan:2007mv}
\begin{equation}
\mu>\mu_\rmcr\equiv\frac{2|\omega|}{(\sqrt{1+\varepsilon}-1)^2}.
\label{eq:mucr}
\end{equation}
This precession behavior of the flavor pendulum in the large
$\mu$ limit represents synchronized oscillations of the bipolar
system with $\Omega_\sync=(1+2\varepsilon^{-1})\omega$. 

\subsection{Bipolar systems with slowly decreasing neutrino density}

\begin{figure*}
\begin{center}                                                
$\begin{array}{@{}c@{\hspace{0.1 in}}c@{}}
\includegraphics*[width=0.48 \textwidth, keepaspectratio]{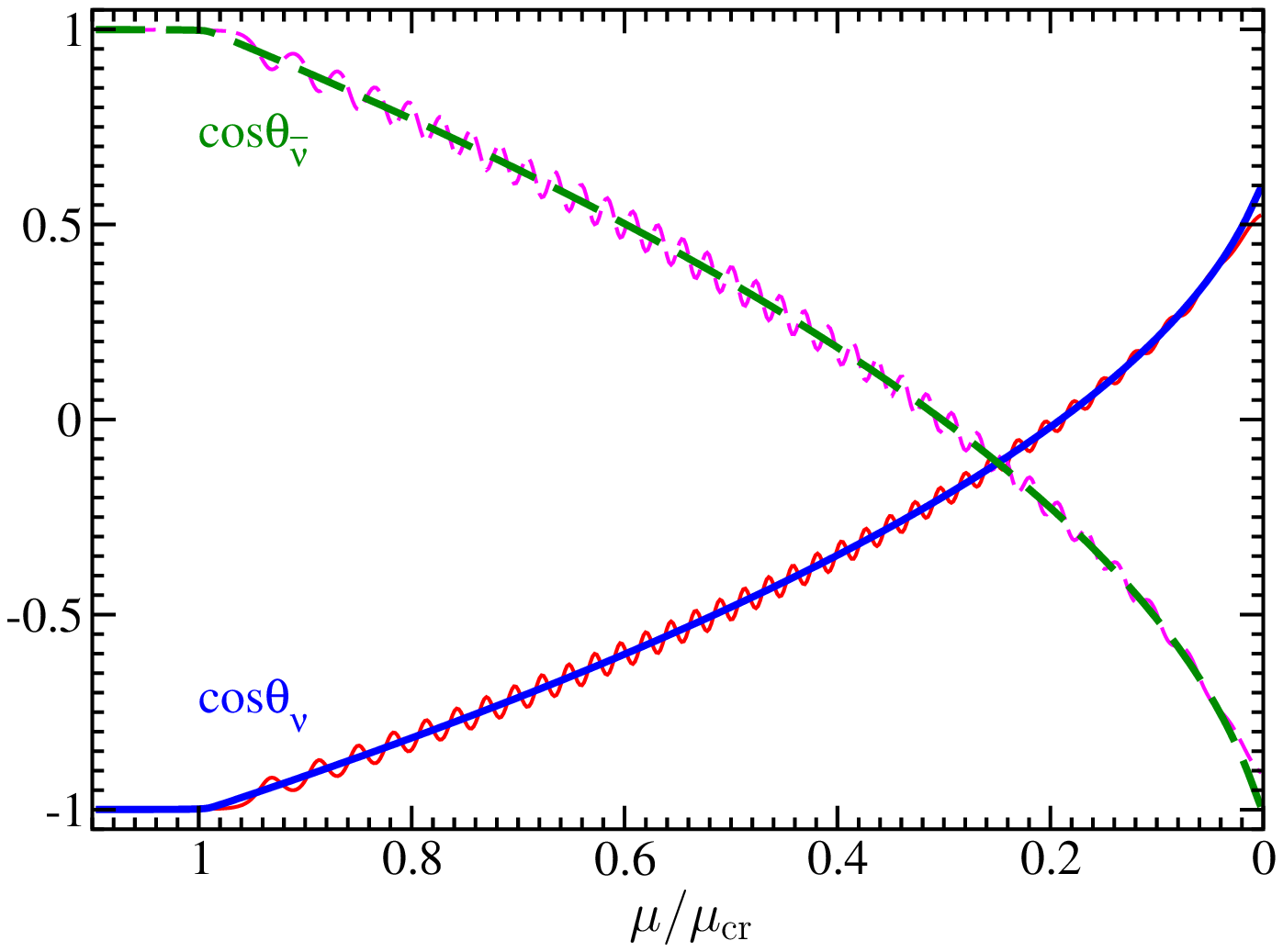} &
\includegraphics*[width=0.48 \textwidth, keepaspectratio]{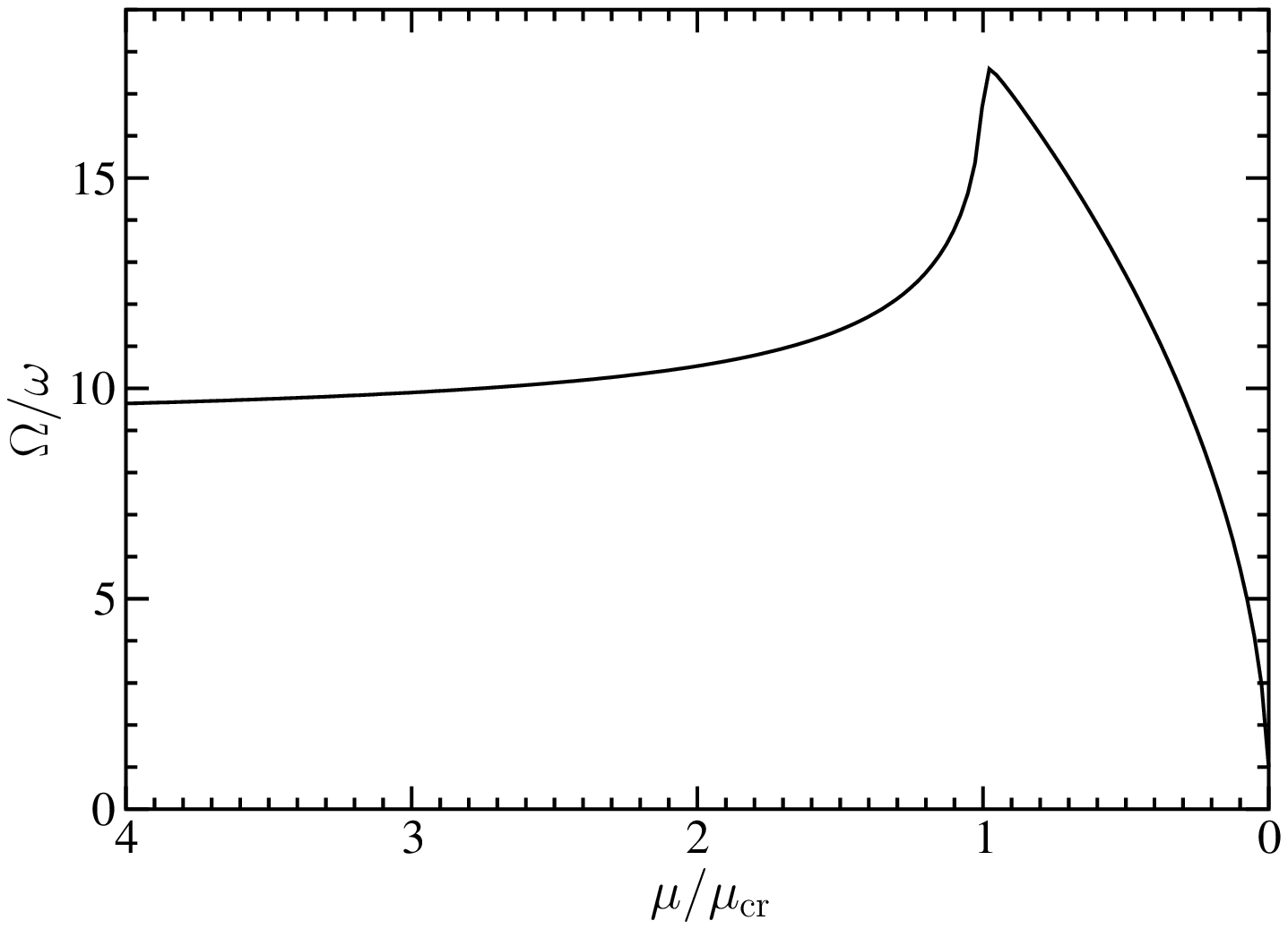} 
\end{array}$
\end{center}
\caption{\label{fig:simple-bipolar}%
Evolution  with decreasing $\mu$ for
a simple bipolar system that  consists initially of
mono-energetic $\nu_e$ and $\bar\nu_e$. In this
example the neutrino excess is $\varepsilon=0.25$ and the
mixing parameters are $\thetav=0.01$ and $\dmsqr<0$. 
The left panel shows the configuration of the polarization vectors,
where $\theta_\nu$ ($\theta_{\bar\nu}$) is the angle between
$\vP_\omega$ ($\vP_{-\omega}$) and $\vB=-\vve_3$. The thick lines
are for an assumed pure precession motion of the flavor pendulum,
 and the thin lines are for a case where
 $\mu$  decreases slowly and linearly with time. At
$\mu\gtrsim\mu_\rmcr$ the flavor pendulum is a ``sleeping top'' and
there are no flavor oscillations. When $\mu<\mu_\rmcr$ the flavor
pendulum experiences precession as well as nutation around the average
precession track. The right panel shows the precession
frequency $\Omega$ of the flavor pendulum assuming that it is
executing  pure precession. 
In this case, the value $\Omega$ approaches $\Omega_\sync$ and $\omega$
in the limits 
$\mu\rightarrow\infty$ and $\mu\rightarrow0$, respectively. The
analytical solution for the pure precession motion of the polarization
vectors can be found in Reference~\cite{Raffelt:2007xt}. This figure
is adapted from
Figures 3(a) and 4 in Reference~\cite{Duan:2007mv}.
Copyright 2007 by the American Physical Society.}
\end{figure*}

Let us now focus on the IH case with $\thetav\ll1$.
Significant neutrino oscillations can occur in this case.
If the neutrino density decreases, the mass $\mu^{-1}$ of the pendulum 
becomes larger. The swing amplitude of the
pendulum that represents a symmetric bipolar system decreases as
$\mu$ decreases. The maximum swing amplitude of this pendulum can be 
found in the adiabatic limit where $\mu$ changes slowly
\cite{Duan:2007mv}. In this case $\vP_\omega$
and $\vP_{-\omega}$ become aligned and antialigned, respectively,
 with $\vB$ (i.e.,
$\nu_e\rightarrow\nu_2$ and $\bar\nu_e\rightarrow\bar\nu_2$) as $\mu$
decreases toward 0.
The asymmetric bipolar system is more
interesting. Neutrino oscillations in this case are synchronized and the flavor
pendulum precesses uniformly in the limit $\mu\gg|\omega|$. If $\mu$
decreases very slowly, the flavor pendulum can still
experience a nearly pure precession motion for any given $\mu$. 
In this case $\vP_\omega$ and $\vP_{-\omega}$
 lie in the same plane with $\vB$ and their directions
can be readily solved for \cite{Duan:2007mv,Raffelt:2007xt}(see
Figure~\ref{fig:simple-bipolar}).  
We note that, assuming $\varepsilon>0$, $\vP_{-\omega}$ becomes
antialigned with $\vB$ (i.e., $\bar\nu_e\rightarrow\bar\nu_2$)
as $\mu\rightarrow0$. Meanwhile the direction of
$\vP_{\omega}$ can be determined from the constancy of
$\vD\cdot\vB$ \cite{Hannestad:2006nj} (see Equation~\myeqref{eq:eom-pendulum},
noting that $\vec{g}\propto\vB$).

\section{Isotropic and homogeneous neutrino gases\label{sec:isotropic-gas}}

\subsection{Static solutions for the neutrino flavor evolution equation
\label{sec:twosols}}

We next consider a homogeneous and isotropic gas that consists of
neutrinos with continuous energy spectra. Let us first seek static
solutions to the EoM for the polarization vectors
(Equation~\myeqref{eq:eom-vP}) in the case where both the matter density 
and the neutrino densities are constant. One possibility is
that the solution is stationary and the $\vP_\omega$'s do not evolve with
time $t$. This is only possible if the ``spin'' $\vP_\omega$ is parallel to
 the corresponding total ``magnetic field''
$\vH_\omega=\omega\vB+\lambda\vL+\mu\vD$. 
For this case, and noting that $\vH_\omega$
does not depend on $\vP_\omega$, we can obtain the following equations
for the total polarization vector $\vD$ \cite{Duan:2007fw}:
\begin{subequations}
\label{eq:align-matt-i}
\begin{align}
D_1&= 
(\lambda\sin2\thetav+\mu D_1)
\int_{-\infty}^\infty\frac{\epsilon_\omega|\vP_\omega|}{|\vH_\omega|}\rmd\omega,
\label{eq:align-matt-1}\\
D_2&= 
\mu D_2
\int_{-\infty}^\infty\frac{\epsilon_\omega|\vP_\omega|}{|\vH_\omega|}\rmd\omega,
\label{eq:align-matt-2}\\
D_3 &= 
\int_{-\infty}^\infty(-\omega+\lambda\cos2\thetav+\mu D_3)
\frac{\epsilon_\omega|\vP_\omega|}{|\vH_\omega|}\rmd\omega,
\label{eq:align-matt-3}
\end{align}
\end{subequations}
where $D_a=\vD\cdot\vve_a$ ($a=1,2,3$), and $\epsilon_\omega=+1$ ($-1$) if
$\vP_\omega$ is aligned (antialigned) with
$\vH_\omega$. 
The stationary solution to the EoM for $\vP_\omega$ can
be found from Equation~\myeqref{eq:align-matt-i} and the alignment
condition.
We note that Equations~\myeqref{eq:align-matt-1} and 
\myeqref{eq:align-matt-2} generally imply that $D_2=0$ if $\lambda\neq0$
which, in turn, 
implies that $\vP_\omega\cdot\vve_2=0$ for any $\omega$.

Equations~\myeqref{eq:align-matt-1} and
\myeqref{eq:align-matt-2} become equivalent when $\lambda=0$,
and $\vD$ is underconstrained by Equation~\myeqref{eq:align-matt-i}
in this case. This is because when $\lambda=0$,
the EoM for $\vP_\omega$
possesses a rotational symmetry about $\vB$.
In other words,
if $\{\vP_\omega(t)|\,\forall\omega\}$ solves
the EoM, then $\{\vP'_\omega(t)|\,\forall\omega\}$
also solves the EoM, where $\vP'_\omega(t)$ is obtained from
$\vP_\omega(t)$ by rotation about $\vB$ by an arbitrary angle
$\phi$. Here $\phi$ is independent of $\omega$ and $t$. This is a
generalization of the rotational symmetry of the flavor pendulum about
the ``gravity'' vector $\vec{g}\propto\vB$. 
Such a symmetry generally implies the existence of a
collective motion and a conservation law. For example, a
translational symmetry of a group
of particles along some direction 
implies the possibility of the collective motion of the
particles in that direction and the conservation of the total
momentum of the particles in the same direction. In
Section~\ref{sec:bipolar} it was shown that this 
rotational symmetry implies the possibility of a
pure precession of the flavor pendulum in which both polarization
vectors of the simple bipolar system precess with the same 
frequency $\Omega$. Therefore, it is natural to seek a static solution
to the EoM with $\lambda=0$, and $\mu$ constant, and in which all
the $\vP_\omega$'s precess about $\vB$ with the same frequency
$\Omega$. In this case all the $\vP_\omega$'s are stationary in a
corotating frame which rotates about $\vB$ with frequency
$\Omega$. In this corotating frame, $\vP_\omega$ is either aligned or
antialigned with $\vtH_\omega=(\omega-\Omega)\vB+\mu\vD$, and 
 the components of $\vD$ in the corotating frame can be found from
equations similar to Equation~\myeqref{eq:align-matt-i}. 
These equations can be recast as the following two simple sum rules
\cite{Raffelt:2007cb}:
\begin{align}
1 &= 
\int_{-\infty}^\infty\frac{\epsilon_\omega |\vP_\omega|}
{\sqrt{[(\Omega-\omega)/\mu+ D_3]^2+ D_\perp^2}}\,\rmd\omega,
\label{eq:sum-1}\\
\Omega &= 
\int_{-\infty}^\infty\frac{\epsilon_\omega \omega |\vP_\omega|}
{\sqrt{[(\Omega-\omega)/\mu+D_3]^2+D_\perp^2}}\,\rmd\omega,
\label{eq:sum-omega}
\end{align}
where $D_\perp$ is the component of $\vD$ that is perpendicular to $\vve_3$.
The rotational symmetry of the system about $\vB$ implies that
 $D_3=-\vD\cdot\vB$ is constant.

\subsection{Adiabatic solutions and the spectral swap/split\label{sec:swap}}

\begin{figure*}
\begin{center}                                                
$\begin{array}{@{}c@{\hspace{0.1 in}}c@{}}
\includegraphics*[width=0.48 \textwidth, keepaspectratio]{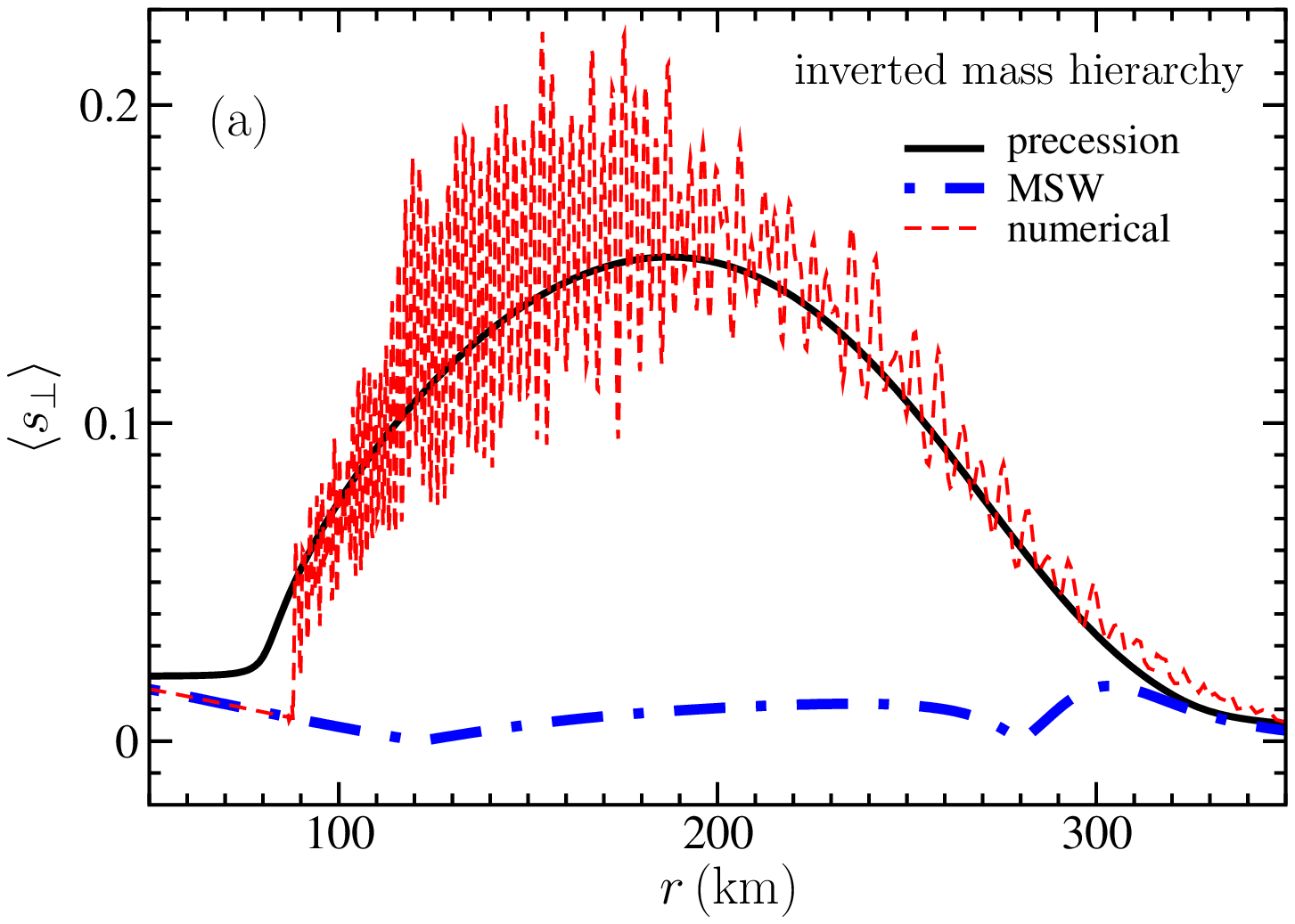} &
\includegraphics*[width=0.48 \textwidth, keepaspectratio]{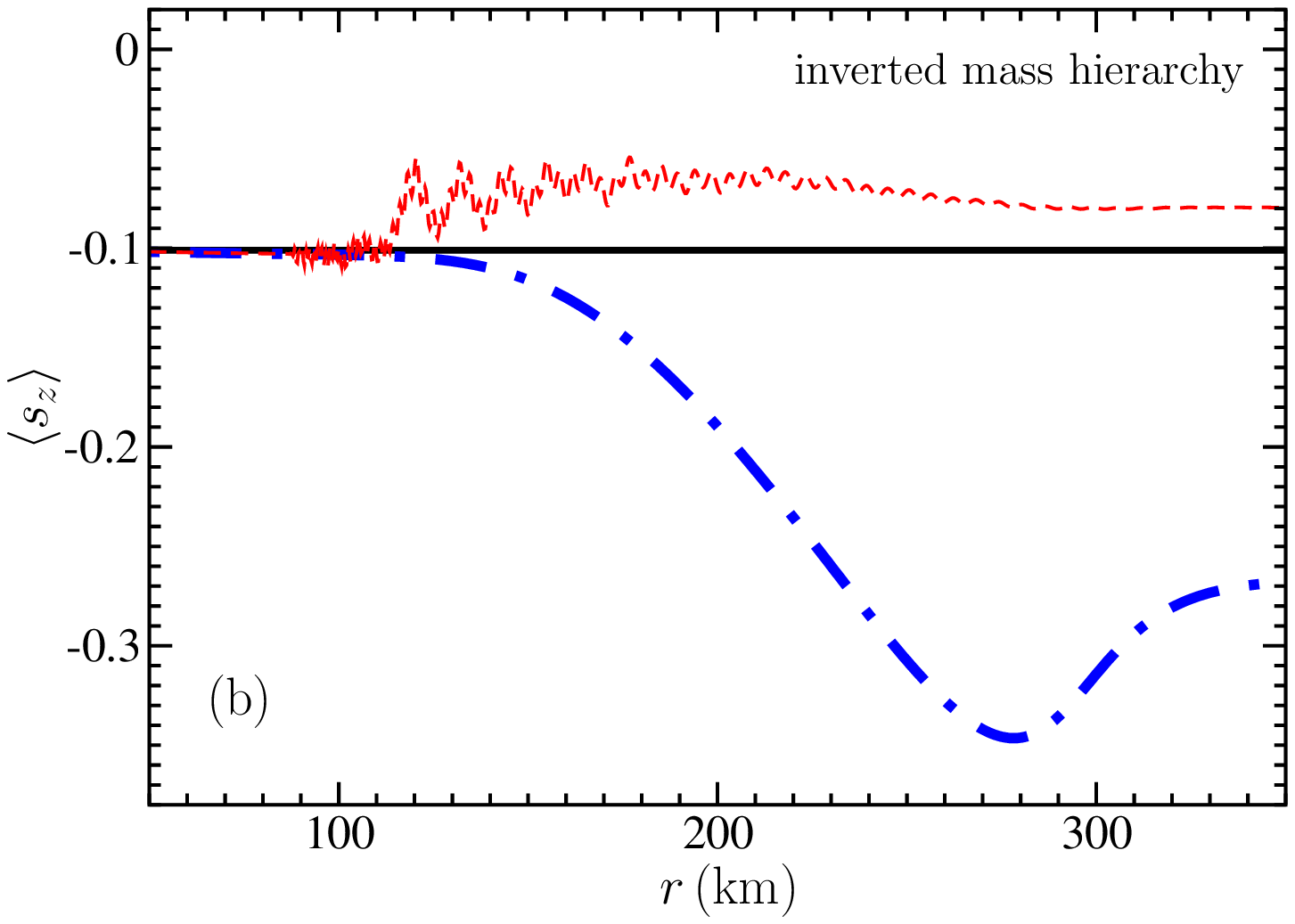} \\
\includegraphics*[width=0.48 \textwidth, keepaspectratio]{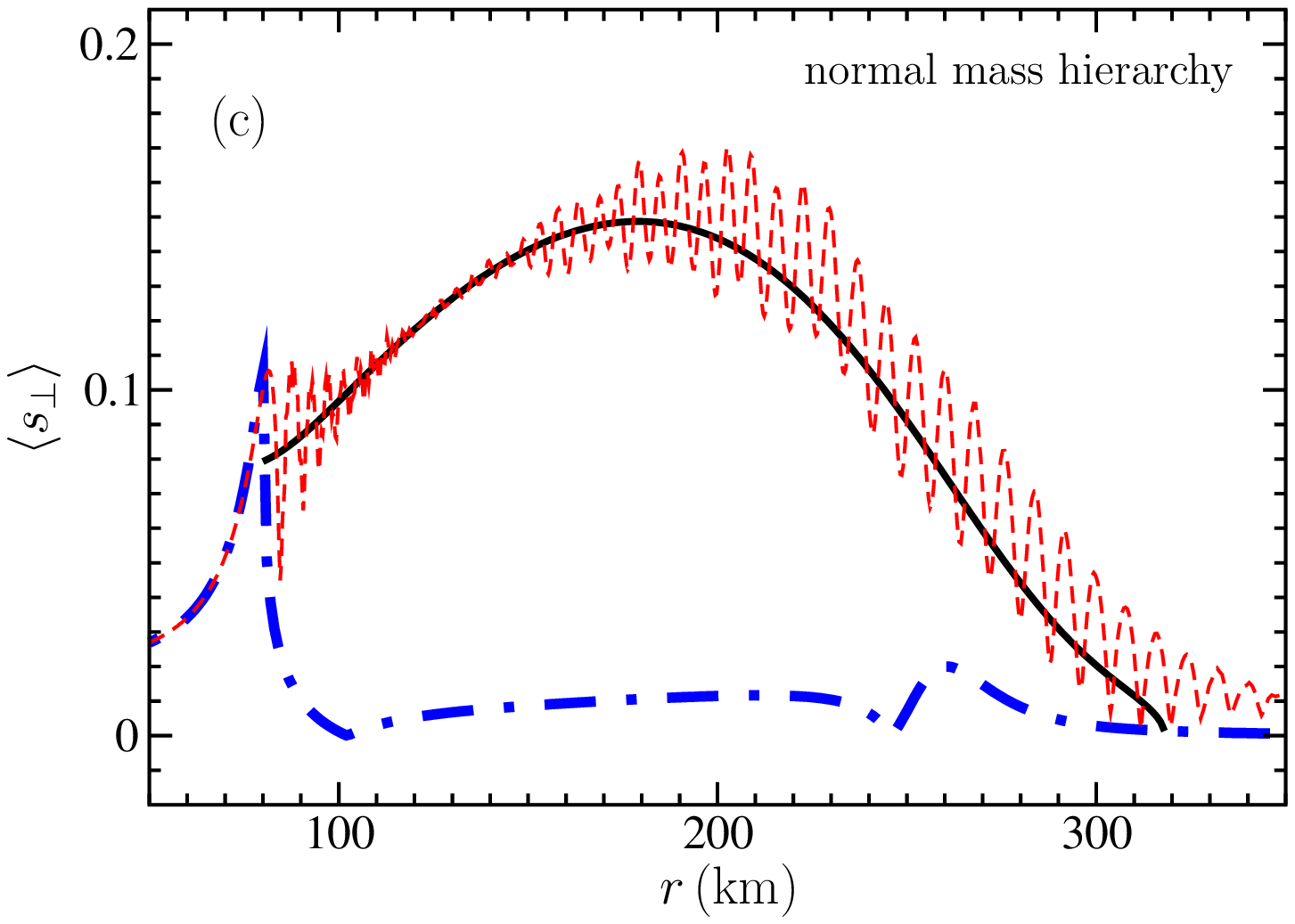} &
\includegraphics*[width=0.48 \textwidth, keepaspectratio]{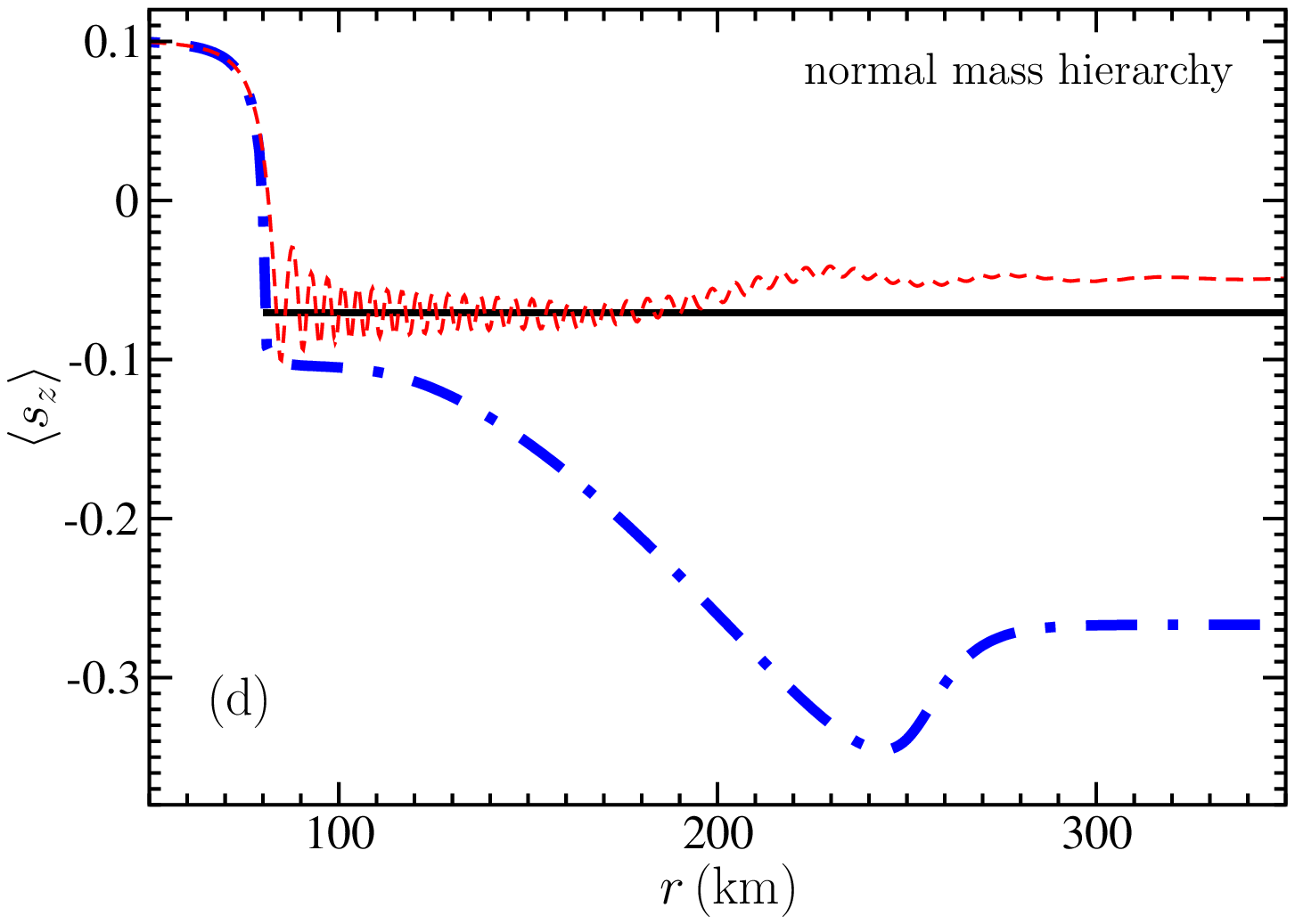}
\end{array}$
\end{center}
\caption{\label{fig:twosols}Comparison of the numerical solution, the
  $\nu$-enhanced MSW solution, and the precession solution in a
  single-angle scheme, where $\thetav=0.1$, 
  $\langle s_\perp\rangle\propto D_\perp$ and 
  $\langle s_z\rangle\propto D_3$. The numerical solution follows the
  MSW solution at first. In the IH case the MSW solution becomes
  unstable at $r\simeq88$ km in this calculation
  and thereafter the numerical solution shows
  oscillations around the precession solution. In the NH
  case the numerical solution follows the MSW solution through the 
  resonance (where $\langle s_z\rangle\simeq0$)
  before it shifts to follow the precession solution track. Figure adapted
  from Figure~1 in Reference~\cite{Duan:2007fw}.
  Copyright 2007 by the American Physical Society.} 
\end{figure*}

If $\lambda$ and $\mu$ vary slowly with time, adiabatic solutions
can be obtained which correspond to the static solutions discussed
above. We shall term these the (adiabatic, $\nu$-enhanced) MSW solution and the
(adiabatic) precession solution.
Figure~\ref{fig:twosols} shows these two adiabatic
solutions for a single-angle scheme together with the
corresponding numerical solutions for Equation~\myeqref{eq:eom-single}. 
In these calculations it is assumed that
the most abundant neutrino species at the neutrino sphere are 
 $\nu_e$ and $\bar\nu_e$, and that $\thetav\ll1$.
For the precession solution the matter
field is ``removed'' using an appropriate corotating frame. 
The flavor pendulum model provides insight into how
 the numerical solution evolves from the MSW solution towards the
precession solution.
In the IH case a large matter density essentially keeps the flavor
pendulum (with $\varepsilon>0$) near its highest position. This
configuration becomes unstable when $\mu<\mu_\rmcr$. In the NH case
the $\nu$-enhanced MSW flavor transformation has the effect of raising
the flavor pendulum to near its highest position, and the
configuration is again unstable when $\mu<\mu_\rmcr$.

\begin{figure*}
\begin{center}
$\begin{array}{@{}c@{\hspace{0.1 in}}c@{}}
\includegraphics*[width=.4\textwidth]{DuanFig6a.eps} &
\includegraphics*[width=.4\textwidth]{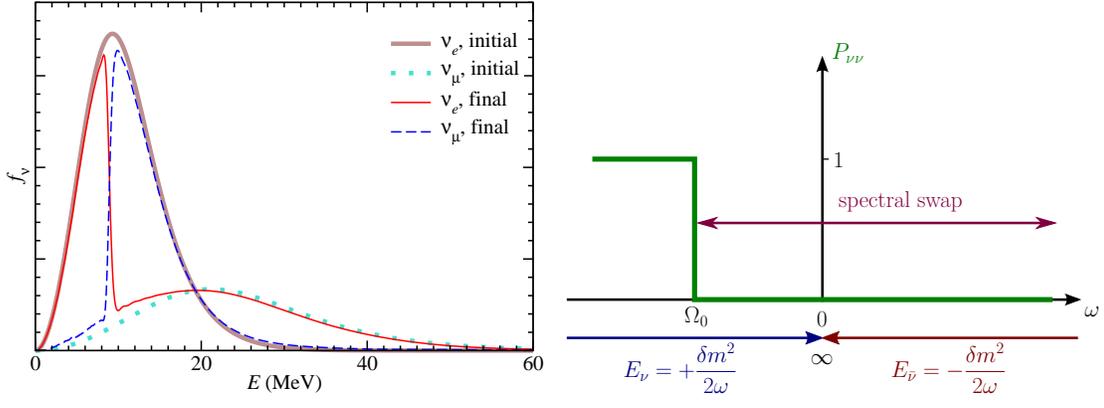} 
\end{array}$
\end{center}
\caption{\label{fig:swap}%
  Illustration of the ``stepwise spectral swap'' phenomenon
  in the two-flavor mixing case with $\dmsqr<0$ (IH) that was discovered in
  Reference~\cite{Duan:2006an}. 
  The left panel shows the stepwise swapping of
  $\nu_e$ and $\nu_\mu$ energy spectra about $E_\spl\simeq9$ MeV in a
  single-angle scheme. The spectra of $\bar\nu_e$ and $\bar\nu_\mu$ are
  nearly fully swapped in this calculation. The right panel shows the
  corresponding survival probability $P_{\nu\nu}$
  which is a step-like function of $\omega$.
  For the NH case the step-like structure of $P_{\nu\nu}(\omega)$ is
  pushed rightward 
  to $\Omega_0>0$. 
  Because $E_\spl=|\frac{\dmsqr}{2\Omega_0}|$  
  splits a neutrino spectrum into two parts with
  different flavors, this phenomenon is also sometimes called ``spectral
  split''. 
  Reprinted with permission
  from H.~Duan, AIP Conf.\ Proc.\ Vol.\ 1182, Page 37, 2009
  (Reference~\cite{Duan:2009rp}).
  Copyright 2009, American Institute of
  Physics.} 
\end{figure*}

Just as in the  conventional adiabatic MSW flavor transformation case
(Figure~\ref{fig:vacuum-MSW}), in the adiabatic precession solution
$\vP_\omega$ follows $\vtH_\omega$ which changes its direction (and magnitude) 
as $\mu$ decreases. This induces
neutrino flavor transformation. In particular, as $\mu\rightarrow0$,
$\vtH_\omega\rightarrow(\omega-\Omega_0)\vB$, 
where $\Omega_0=\Omega(\mu=0)$. 
This means that the adiabatic collective precession mode converts the
initial $\nu_e$ into the mass state $|\nu_1\rangle$ or $|\nu_2\rangle$
depending on whether $\omega$ is smaller or larger than $\Omega_0$
\cite{Duan:2006an}. 
This phenomenon,  
the ``stepwise spectral swap'' or ``spectral split'',
is most dramatic when $\thetav\ll1$
 (see Figure~\ref{fig:swap}). The swap/split energy
$E_\spl=|\frac{\dmsqr}{2\Omega_0}|$ can be determined from the
constancy of $\vD\cdot\vB$ \cite{Raffelt:2007cb}.

\subsection{Precession solution in the three-flavor
  mixing scenario\label{sec:three-flavor}}

The neutrino polarization vector defined in Equation~\myeqref{eq:vP} can be
 generalized easily to the three-flavor mixing scenario by replacing 
the Pauli matrices with the Gell-Mann matrices $\Lambda_a$
($a=1,2,\ldots,8$) \cite{Kim:1987bv,Dasgupta:2007ws}. However, because an
eight-dimensional polarization vector, or Bloch vector, cannot be as easily
visualized as its three-dimensional counterpart, we will discuss the
collective precession mode by using the matrix
formalism. To this end, we define the polarization matrix
$\sfP_\omega=\frac{1}{2}\sum_{a=1}^8(P_{\omega,a}\Lambda_a)$ where
$P_{\omega,a}$ is the $a$th component of the Bloch vector
$\vP_\omega$. The polarization matrix obeys the EoM
\begin{equation}
\rmi\dot{\sfP}_\omega =
[\omega_\rmL\sfB_\rmL+\omega_\rmH\sfB_\rmH + \mu\sfD,\, \sfP_\omega],
\label{eq:eom-P}
\end{equation}
where 
$\sfD = \int_{-\infty}^\infty\sfP_\omega\rmd\omega$
is the total polarization matrix.
In Equation~\myeqref{eq:eom-P},
$\omega_\rmL=\pm\frac{\delta  m^2}{2E}$ and
$\sfB_\rmL=-\frac{1}{2}\Lambda_3$ (in the mass basis) 
correspond to the small mass splitting which we define as
$\delta m^2\equiv m_2^2-m_1^2\simeq\dmsqr_\odot$.
Also in Equation~\myeqref{eq:eom-P}, $\omega_\rmH=\omega=\pm\frac{\dmsqr}{2E}$
and $\sfB_\rmH=-\frac{1}{\sqrt{3}}\Lambda_8$ (in the mass basis)
correspond to the large
mass splitting which we define as
$\dmsqr=m_3^2-\frac{1}{2}(m_1^2+m_2^2)\simeq\pm\dmsqr_\mathrm{atm}$.
For simplicity we have ignored the matter field since it can be removed
by employing the corotating frame technique \cite{Duan:2008za}.
 
The static precession solution (with constant $\mu$) can 
be obtained by assuming that  
$\{\stP_\omega,\Omega_\rmL,\Omega_\rmH|\,\forall\omega\}$ solves
equation 
\begin{equation}
[\stH_\omega,\,\stP_\omega] = 
[(\omega_\rmL-\Omega_\rmL)\sfB_\rmL+(\omega_\rmH-\Omega_\rmH)\sfB_\rmH
+\mu\stD,\, \stP_\omega] = 0,
\label{eq:precession-ansatz}
\end{equation}
where $\stP_\omega$, $\Omega_\rmL$ and $\Omega_\rmH$ are constant, and
$\stD=\int_{-\infty}^\infty\stP_\omega\rmd\omega$. In
the two-flavor mixing scenario ($\sfB_\rmL=0$),
Equation~\myeqref{eq:precession-ansatz} corresponds to the condition that
the ``spin'' $\vec{\tilde{P}}_\omega$ is stationary in a corotating frame
and is parallel to the total ``magnetic field'' 
$\vtH_\omega$ in this reference frame. Equation~\myeqref{eq:precession-ansatz} is
called the precession 
ansatz \cite{Duan:2008za} because it implies a static precession
solution to the EoM (Equation~\myeqref{eq:eom-P}). This solution, 
$\{\sfP_\omega(t),\Omega_\rmL,\Omega_\rmH|\,\forall\omega\}$, can be
written as
\begin{equation}
\sfP_\omega(t)=\exp[-\rmi(\Omega_\rmL\sfB_\rmL+\Omega_\rmH\sfB_\rmH)t]
\,\stP_\omega\,\exp[\rmi(\Omega_\rmL\sfB_\rmL+\Omega_\rmH\sfB_\rmH)t].
\label{eq:prec-sol3}
\end{equation}

\begin{figure}
\begin{center}
$\begin{array}{@{}c@{\hspace{0.1 in}}c@{}}
\includegraphics*[width=0.48 \textwidth,
  keepaspectratio]{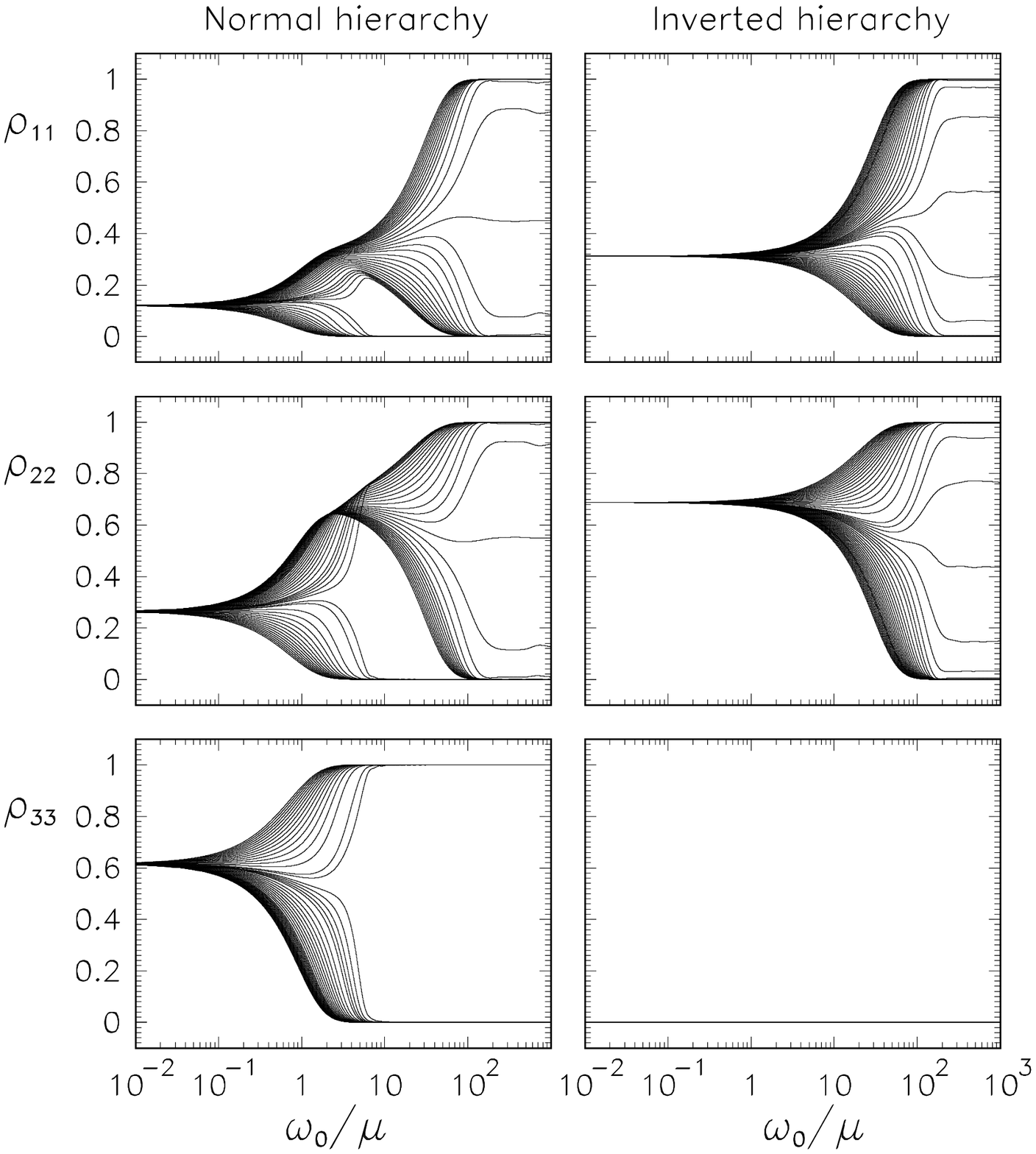} &
\includegraphics*[width=0.48 \textwidth,
  keepaspectratio]{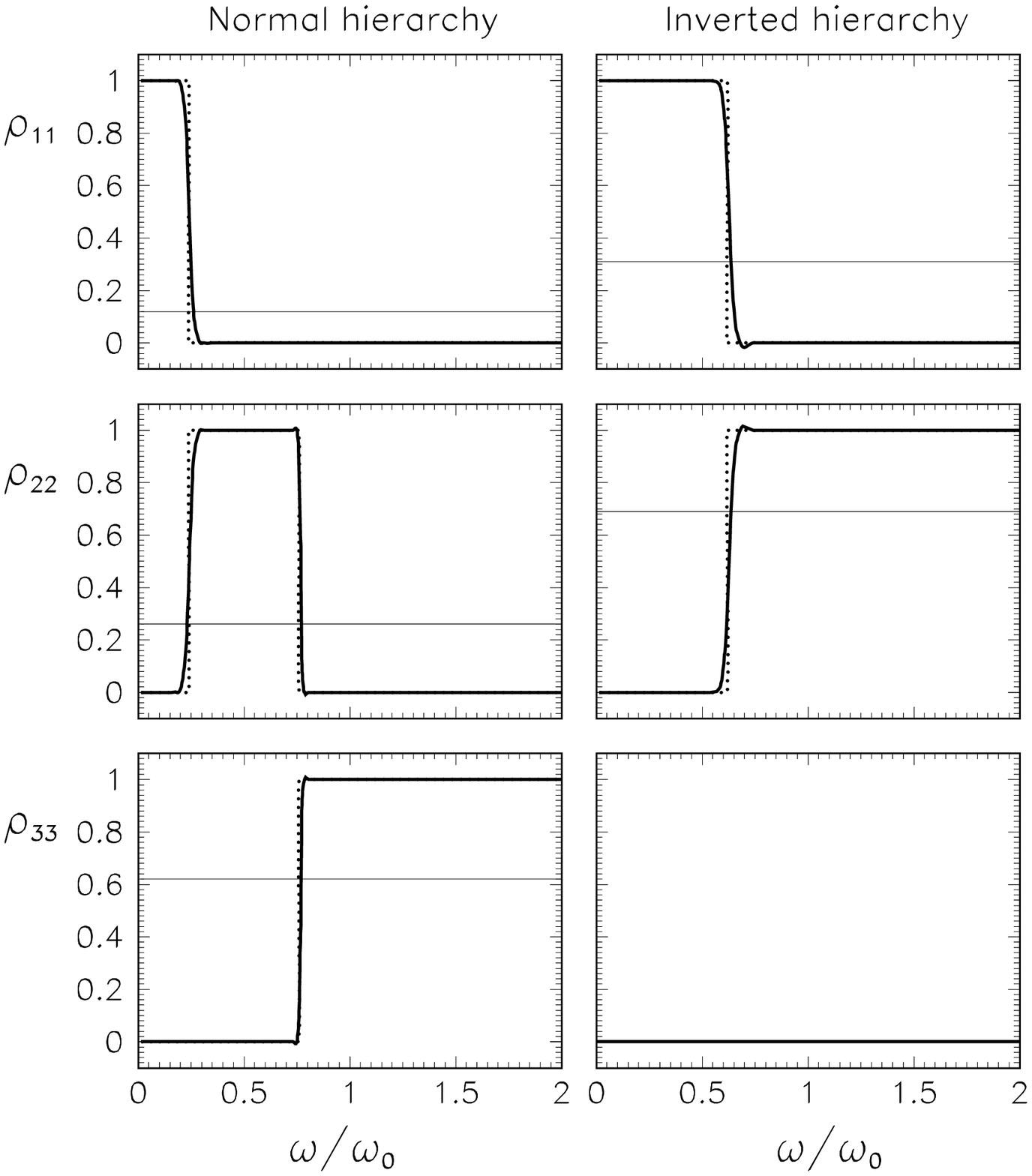}
\end{array}$
\end{center}
\caption{\label{fig:reduction}%
  The evolution of the diagonal elements of the neutrino density
  matrices in the mass basis during the   hierarchical formation of
  spectral  swaps/splits in a neutrino gas.
  Initially 
  $\sfP_\omega$, and the density matrix $\rho_\omega$, is the same across
  the whole energy spectrum $\omega\in[0,2\omega_0]$
  (the thin lines in the panels on the right). 
  When $\mu\gg\omega_0$,
  all neutrinos are in the heaviest eigenstate (i.e.,
  the state with the largest eigenvalue) of
  $\stH_\omega\simeq\mu\stD$.
  For the NH case two spectral swaps/splits form hierarchically at
  $\mu\sim\omega_0$ and $\mu\sim\alpha\omega_0$, respectively. Curves
  in the left-hand panels give $\rho_\omega$ with various values of
  $\omega$. At 
  $\omega_0\gg\mu\gg\alpha\omega_0$ the heaviest eigenstate of
  $\stH\simeq(\omega-\Omega_{\rmH,0})\sfB_\rmH+\mu\stD$ is
  $|\nu_3\rangle$ for neutrinos with
  $\omega>\Omega_{\rmH,0}$. Subsequently these neutrinos no longer
  participate in collective oscillations.
  At $\mu\ll\alpha\omega_0$ the remaining neutrinos are in
  $|\nu_2\rangle$ or $|\nu_1\rangle$, depending on whether
  $\omega_\rmL=\alpha\omega$ is larger or smaller 
  than $\Omega_{\rmL,0}$. In either case the state corresponds to
  the heaviest eigenstate of
  $\stH_\omega\simeq(\omega_\rmL-\Omega_{\rmL,0})\sfB_\rmL$ (ignoring the
  decoupled state $|\nu_3\rangle$). 
  This leads to two spectral swaps/splits in the
  final neutrino energy spectra (the thick lines in the NH
  column in the right-hand panels). 
  In the IH case, initially the $\rho_{33}$ are set to  0 for all
  $\omega$. As a 
  result $|\nu_3\rangle$ is completely decoupled from the EoM
  and no spectral swap/split forms at $\mu\sim\omega_0$.
  The dotted lines in the right-hand panels are computed by
  using the constancy of   $\Tr(\sfD\cdot\Lambda_3)$ and
  $\Tr(\sfD\cdot\Lambda_8)$ and 
  by assuming that the spectral swaps/splits are  infinitely sharp.
  Reprinted figures with permission from 
  \href{http://prd.aps.org/abstract/PRD/v77/i11/e113007}{B.~Dasgupta et al, 
  Phys.\ Rev.\ D, Vol.\ 77, 113007, 2008} 
  (Reference~\cite{Dasgupta:2008cd}). 
  Copyright 2008 by the American Physical Society.}
\end{figure}

Because $\stP_\omega$ and $\stH_\omega$ commute, they
can be simultaneously diagonalized by a unitary matrix
$\sfX$. Therefore, we have
$\sfX\stH_\omega\sfX^\dagger=\diag[\tlh_{\omega,1},\tlh_{\omega,2},\tlh_{\omega,3}]$,
where $\tlh_{\omega,1}<\tlh_{\omega,2}<\tlh_{\omega,3}$, and
$\sfX\stP_\omega\sfX^\dagger=\diag[\tlp_{\omega,1},\tlp_{\omega,2},
-(\tlp_{\omega,1}+\tlp_{\omega,2})]$.
When $\mu$  varies slowly with $t$, the adiabatic precession solution
can be obtained from the static precession solutions (with different
$\mu$) by using the adiabatic ansatz \cite{Duan:2008za}
\begin{equation}
\frac{\partial}{\partial\mu}\tlp_{\omega,1}
=\frac{\partial}{\partial\mu}\tlp_{\omega,2}=0.
\label{eq:adiabatic-ansatz}
\end{equation}
In the two-flavor mixing scenario, the adiabatic ansatz corresponds
to the assumption that $\vP_\omega$ remains parallel to $\vtH_\omega$.
If neutrinos follows the adiabatic precession solution,
then as $\mu\rightarrow0$, 
$\stP_\omega$ becomes a diagonal matrix in the mass basis  (see
Equations~\myeqref{eq:precession-ansatz}). The 
diagonal elements of $\sfP_\omega|_{\mu=0}=\stP_\omega|_{\mu=0}$ are
$\tlp_{\omega,1}$, 
$\tlp_{\omega,2}$ and $-(\tlp_{\omega,1}+\tlp_{\omega,2})$, and these
elements have the same order of appearance as
do the diagonal elements of
$\stH_\omega|_{\mu=0}=(\omega_\rmL-\Omega_\rmL)\sfB_\rmL+(\omega_\rmH-\Omega_\rmH)\sfB_\rmH$. For
example, in the mass basis
we will have $\sfP_\omega|_{\mu=0}=\diag[\tlp_{\omega,2},\tlp_{\omega,1},
-(\tlp_{\omega,1}+\tlp_{\omega,2})]$ if
$\stH_\omega|_{\mu=0}=\diag[\tlh_{\omega,2},\tlh_{\omega,1},\tlh_{\omega,3}]$,
where $\tlh_{\omega,1}<\tlh_{\omega,2}<\tlh_{\omega,3}$. The fact that
$\sfP_\omega|_{\mu=0}$ is diagonal 
in the mass basis implies that there can be
 multiple spectral swaps/splits in the final neutrino energy
spectra. Because $\alpha=\frac{\delta m^2}{|\dmsqr|}\ll1$,
these spectral swaps/splits form hierarchically and
appear at different neutrino densities (see Figure~\ref{fig:reduction}).

\section{Anisotropic and/or inhomogeneous neutrino gases
\label{sec:anisotropic-gas}}

\subsection{Kinematic decoherence of collective neutrino
  oscillations\label{sec:decoherence}}

The oscillations of neutrinos with different momenta
can become out of phase (i.e., the breakdown of collective
oscillations). This has sometimes been called
``kinematic decoherence''. 
Of course, this is not to be confused with quantum decoherence, 
which can be induced by any neutrino scattering process that changes
neutrino momentum.
In a homogeneous and isotropic neutrino gas the condition for the kinematic
decoherence is $\Delta \omega \gg \mu$. In this limit the
coupling among ``spins'' is not strong enough to maintain a collective motion
and $\vP_\omega$ will precess about $\vB$ with vacuum
oscillation frequency $\omega$.
In Section~\ref{sec:synchronization} we have seen that, 
when $\Delta \omega \ll \mu$,
synchronized neutrino oscillations do not decohere kinematically, and
$|\vD|$ is approximately constant because of energy conservation. 
Numerical simulations suggest that bipolar neutrino
oscillations are also stable
\cite{Kostelecky:1993yt,Kostelecky:1993ys}, although as yet
we know of no conservation law which could explain this.

If collective neutrino oscillations also exist in an anisotropic
environment, then the wavefronts of the oscillation waves of the
neutrinos must coincide with one another. In the neutrino bulb model
these wavefronts are spheres that co-center with the PNS. 
However, in this model a neutrino propagating along a non-radial
trajectory would 
travel a distance longer than that of a radially propagating
neutrino \cite{Qian:1994wh}. In other words, a non-radially
propagating neutrino appears 
to have a larger oscillation frequency $\omega'=\omega/\cos\vartheta$
along $r$ 
(see Equation~\myeqref{eq:eom-bulb}). This will enlarge $\Delta\omega$
and would require stronger neutrino self-coupling to
maintain collective oscillations among neutrinos propagating along
different trajectories. 
Note that the ``magnetic field''
$\vH_{\nu\nu,\vartheta}(r)$ generated by other ``spins'' is trajectory
dependent. This adds another potential source of  kinematic decoherence. 

Reference~\cite{Raffelt:2007yz} pointed out that
a neutrino gas  with an initially symmetric bipolar
configuration (i.e., $\vP_{\nu}=-\vP_{\bar\nu}$) can experience quick
kinematic decoherence even in the presence of a small anisotropy,
and in this case both $|\vP_\nu|$ and $|\vP_{\bar\nu}|$
evolve towards 0. 
Later it was found that an asymmetric bipolar
neutrino system (i.e., $|\vP_\nu|=(1+\varepsilon)|\vP_{\bar\nu}|$ with
$\varepsilon\neq0$) may nor may not experience kinematic decoherence
depending on the value of $\varepsilon$
\cite{EstebanPretel:2007ec}. For typical choices of other 
parameters in a neutrino bulb model, kinematic decoherence is suppressed if
$\varepsilon\gtrsim0.3$.  
 
In the above discussion we have ignored ordinary matter. In the
presence of a large matter density, $\vP_\omega$ tends to precess
around $-\vL$
with frequency $\omega'=-\lambda+\omega\cos2\thetav$. In the isotropic  
environment $\omega'\rightarrow\omega\cos2\thetav$ in the corotating
frame that rotates about $\vL$ with frequency
$\lambda$. Therefore, in this case
when neutrinos experience collective oscillations
ordinary matter can be ignored. In an anisotropic environment, such as the
neutrino bulb model, we will have frequency
$\omega'=(-\lambda+\omega\cos2\thetav)/\cos\vartheta$
along $r$. Clearly, it is
not possible to remove the matter effect for all neutrino
trajectories. This implies that
collective neutrino oscillations would not exist in a region where
the net number density of electrons is much larger than that of
neutrinos \cite{EstebanPretel:2008ni}.

\subsection{Precession mode in the anisotropic environment}

In Section~\ref{sec:bipolar} we have seen that with $\varepsilon\neq0$
the flavor pendulum possesses an internal spin. The existence of this
internal spin makes it possible for the flavor pendulum to experience
simultaneously a precession motion and a nutation motion. These two
kinds of motion of the flavor pendulum correspond to the precession
mode and the bipolar mode of 
neutrino oscillations. The findings in
References~\cite{Raffelt:2007yz} and \cite{EstebanPretel:2007ec}
suggest that bipolar neutrino oscillations, which are the most
prominent when $\varepsilon$ is small, becomes non-collective in the
anisotropic environment. These findings also suggest that the
precession mode, which becomes important for cases with 
sufficiently large $\varepsilon$,
can remain collective in the anisotropic environment. This is
partially confirmed by the fact that neutrino oscillations 
calculated in the
single-angle scheme and in the multi-angle scheme possess common
features, such as the spectral swap/split phenomenon.
This phenomenon, as explained with the single-angle supernova scheme
or in homogeneous and isotropic environments,
results from neutrino oscillations in the
collective precession mode.

\begin{figure}
\begin{center}
\includegraphics*[width=\textwidth, keepaspectratio]{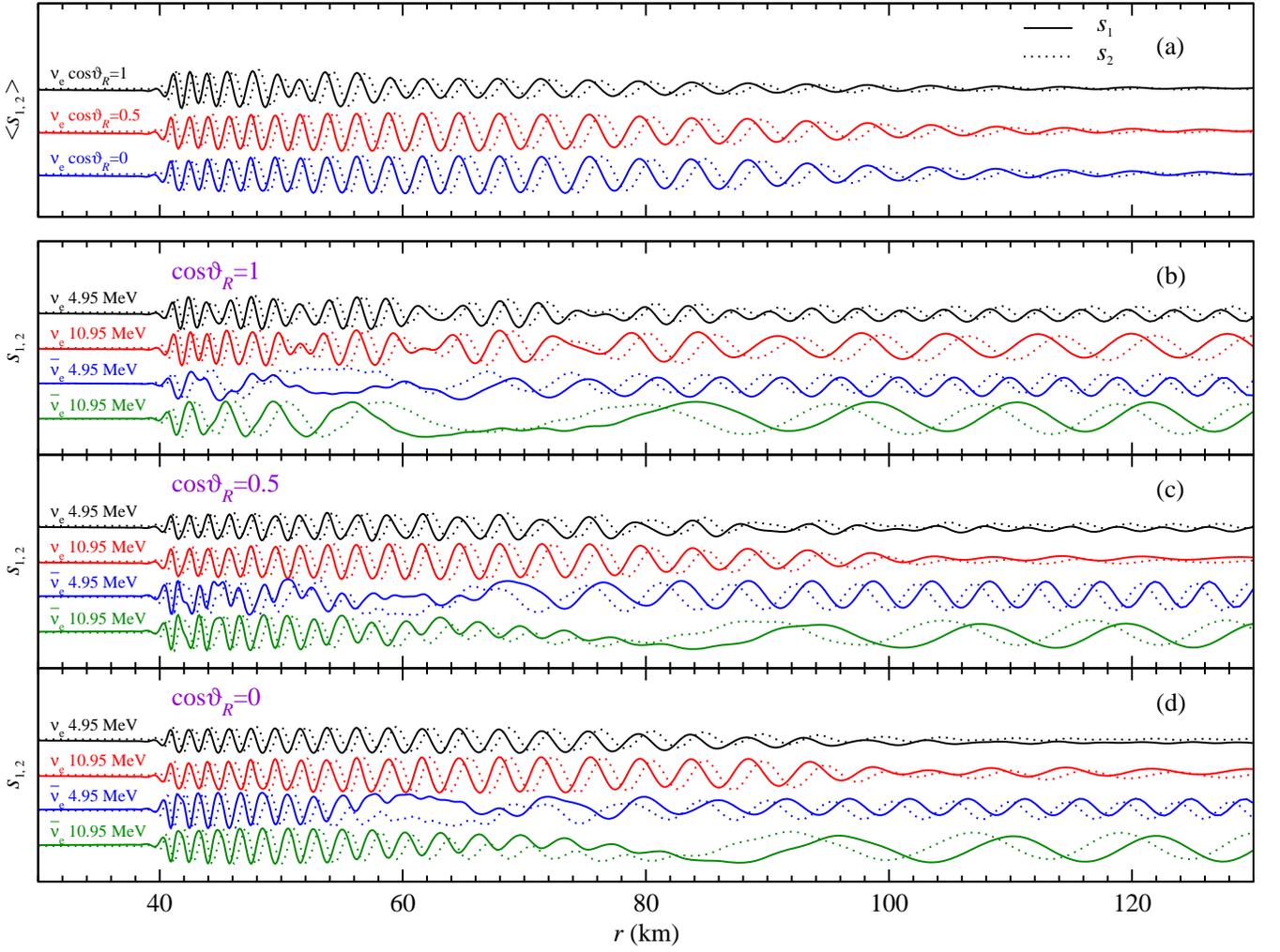}
\end{center}
\caption{\label{fig:spinwave}%
The collective precession of $\vP_{\omega,\vartheta}$ in a neutrino
bulb model with $\thetav\ll1$ and the inverted mass hierarchy
($\dmsqr<0$). The bottom three panels show that for most neutrinos
the ``spin components''
$s_1\propto\vP_{\omega,\vartheta}\cdot\vve_1$ and 
$s_2\propto\vP_{\omega,\vartheta}\cdot\vve_2$ 
oscillate in phase right after collective neutrino oscillations begin
(for this case, at $r\simeq40$ km). 
The in-phase oscillations of $s_1$ and $s_2$ and the
constant relative phase of these components 
imply the collective precession of the
polarization vectors. The neutrinos or antineutrinos that drop out of
the collective precession mode at smaller radii
 are those with vacuum oscillation frequencies
farther from the collective oscillation frequency and those propagating
along trajectories with larger $\cos\vartheta_R$. 
The top panel shows energy-averaged components, $\langle s_1\rangle$
and $\langle s_2\rangle$.
This panel shows that the collective precession mode stands out
when non-collective neutrino oscillations average to zero.
Figure adapted from Figure~1 in Reference~\cite{Duan:2008fd}.
Copyright 2009 by the Institute of Physics.}
\end{figure}

The EoM of the polarization vectors in a
stationary environment can be
written as \cite{Sigl:1992fn,Strack:2005ux,Cardall:2007zw}
\begin{equation}
\bhp\bcdot\boldsymbol{\nabla}\vP_{\omega,\bhp}(\bfx)=
[\omega\vB+\vH_{\nu\nu,\bhp}(\bfx)]\times\vP_{\omega,\bhp}(\bfx),
\label{eq:eom-general}
\end{equation}
where we have assumed $\lambda=0$ for now,
all the polarization vectors are normalized by 
$n_\nu(\bfx_0)$, the number density of neutrino species
$\nu$ at location $\bfx_0$, and 
\begin{equation}
\vH_{\nu\nu,\bhp}(\bfx)=\sqrt{2}\GF n_\nu(\bfx_0)
\int_{-\infty}^\infty\rmd\omega'
\int\rmd^2\bhp'(1-\bhp\bcdot\bhp')\vP_{\omega',\bhp'}(\bfx).
\end{equation}
Like the isotropic-neutrino-gas case, Equation~\myeqref{eq:eom-general}
also exhibits rotational symmetry about $\vB$. If
$\{\vP_{\omega,\bhp}(\bfx)|\,\forall\omega,\bhp\}$ solves 
the EoM, then $\{\vP'_{\omega,\bhp}(\bfx)|\,\forall\omega,\bhp\}$
also solves the EoM, where $\vP'_{\omega,\bhp}(\bfx)$ is obtained from
$\vP_{\omega,\bhp}(\bfx)$ by rotation about $\vB$ by an arbitrary angle
$\phi$, and where $\phi$ is independent of $\omega$, $\bhp$, and
$\bfx$. This symmetry leads to the conservation law \cite{Duan:2008fd}
\begin{equation}
\boldsymbol{\nabla}\bcdot\left[
\int_{-\infty}^\infty\rmd\omega\int\rmd^2\bhp\,
(\vP_{\omega,\bhp}\cdot\vB)\bhp\right]=0.
\end{equation} 
The rotational symmetry of the EoM about $\vB$ can lead to a
collective precession mode for
neutrino oscillations, as shown in Figure~\ref{fig:spinwave}, 
even in an anisotropic environment.

Let $\bfK$ be the wave vector of the collective neutrino
oscillation wave in a stationary environment. If $\vP_{\omega,\bhp}$
experiences pure precession, then this vector must precess about $\vB$ with
frequency $\bhp\bcdot\bfK$ (in flavor space) as 
the corresponding neutrino  propagates along its
word line (in coordinate space). Similar to the isotropic-neutrino-gas
case, this means that  $\vP_{\omega,\bhp}$
is parallel to 
$\vtH_{\omega,\bhp}=(\omega-\bhp\bcdot\bfK)\vB+\vH_{\nu\nu,\bhp}$. 
Therefore, the corresponding neutrino or antineutrino must be in a
mass eigenstate if \cite{Duan:2008fd}
\begin{equation}
|\omega-\bhp\bcdot\bfK|\gg |\vH_{\nu\nu,\bhp}|.
\label{eq:coll-criterion}
\end{equation}
Equation~\myeqref{eq:coll-criterion} gives the criterion for when
neutrinos or antineutrinos drop out of the collective precession
mode and begin to oscillate incoherently with respect to
other neutrinos.
The neutrinos or antineutrinos which drop out in this way have
oscillation frequencies $\omega$ so different from the collective oscillation
frequency that neutrino self-interaction is not strong enough to
maintain the corresponding polarization vectors in collective precession. 

With the replacements
$\omega\rightarrow(\omega\cos2\thetav-\lambda)$ and
$\vB\rightarrow-\vL$
the above discussion also applies to the case with a large $\lambda$.
With these replacements, Equation~\myeqref{eq:coll-criterion}
 indicates that collective precession is
indeed suppressed when 
the number density of electrons is much larger than that of
neutrinos.

\section{Collective neutrino oscillations in supernovae
\label{sec:supernovae}}

\subsection{Neutrino oscillation regimes}

\begin{figure*}
\begin{center}                                                
$\begin{array}{@{}c@{\hspace{0.1 in}}c@{}}
\includegraphics*[width=0.48 \textwidth, keepaspectratio]{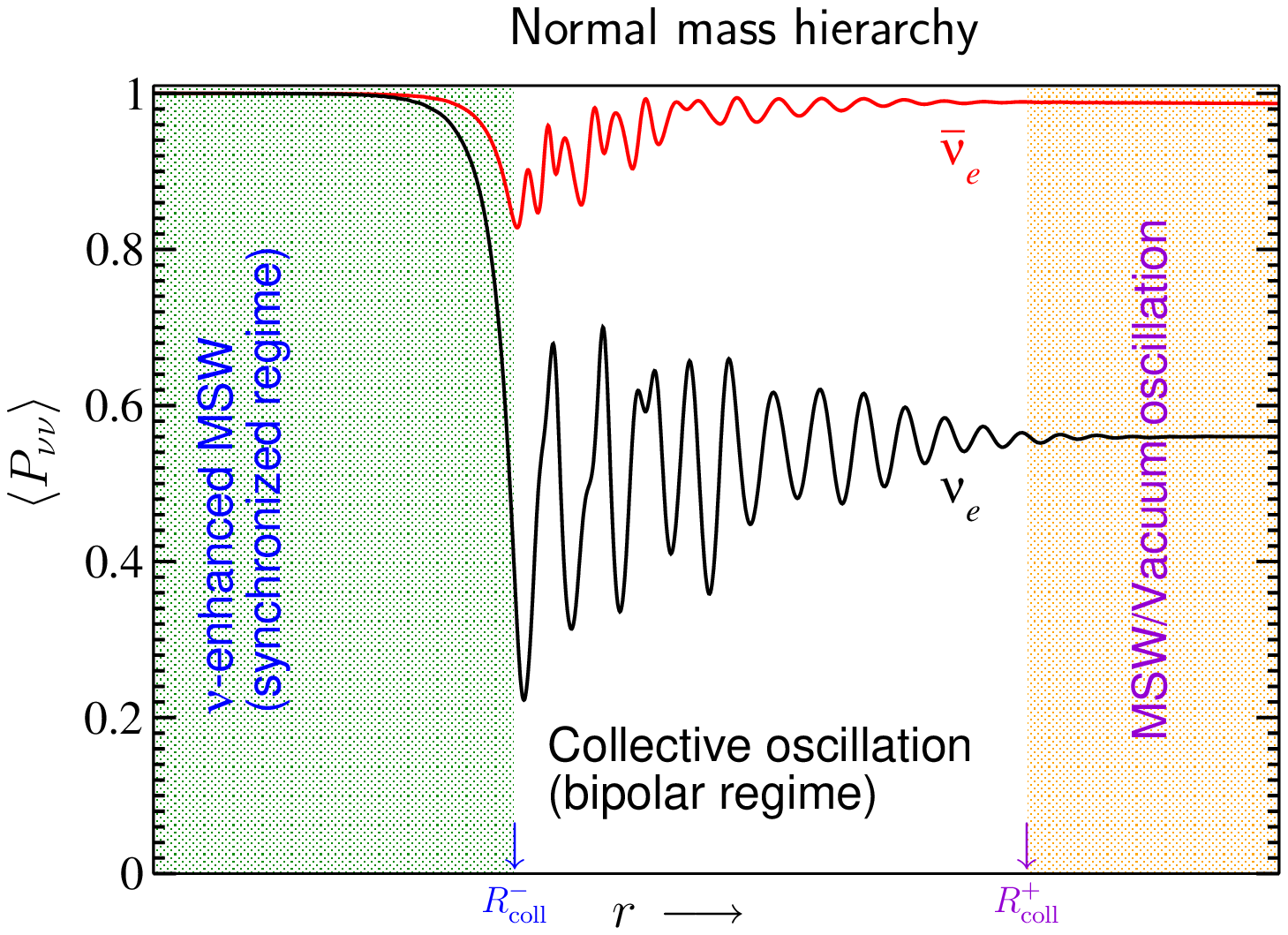} &
\includegraphics*[width=0.48 \textwidth, keepaspectratio]{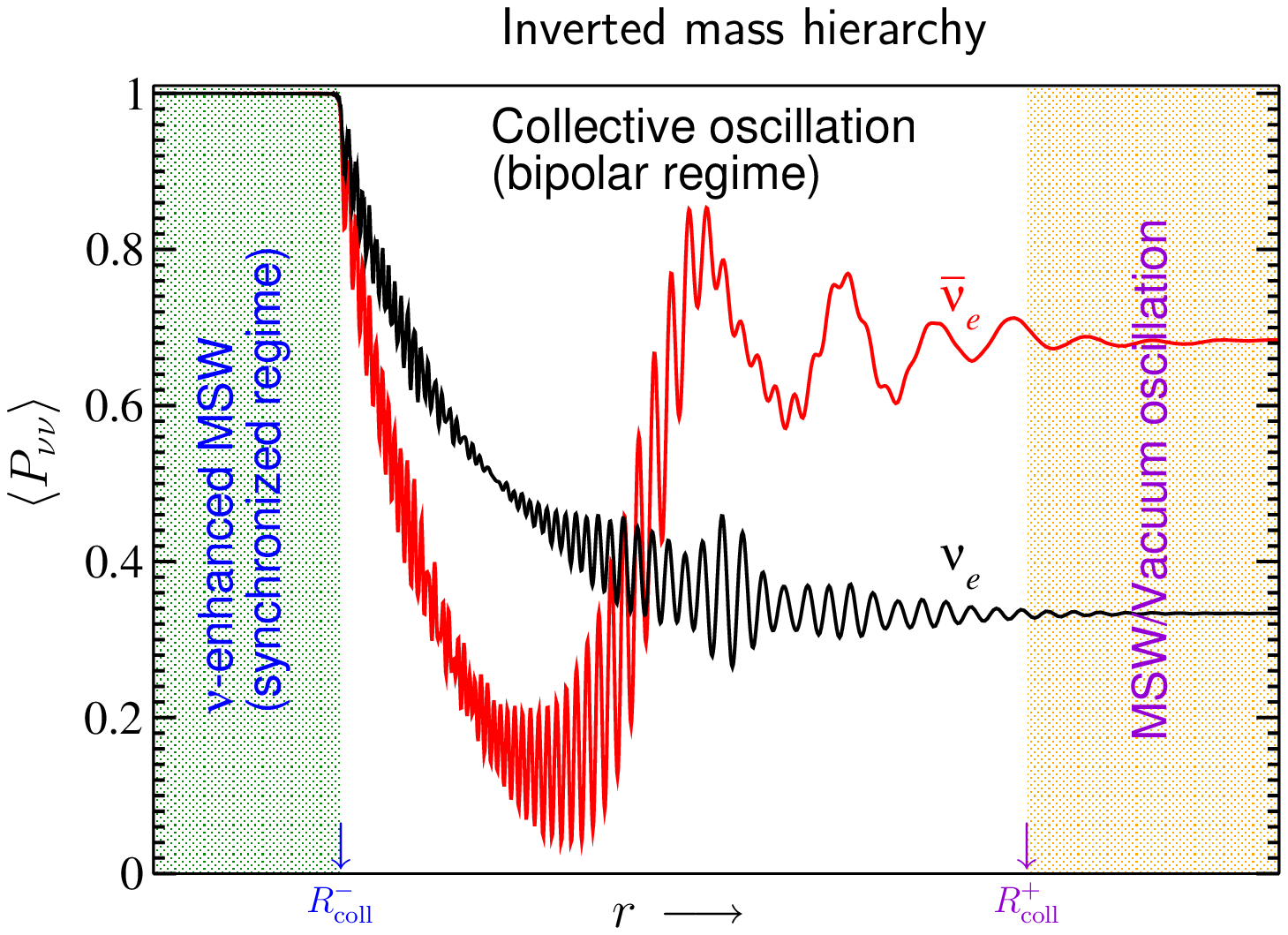}
\end{array}$
\end{center}
\caption{\label{fig:Pnunu}Schematic plot of the neutrino oscillation
  regimes in the core collapse supernova environment in the two-flavor
  mixing scenario with 
  $|\dmsqr|\simeq\dmsqr_\mathrm{atm}$. Near the PNS, and in the  
  ``synchronized regime'' ($r\lesssim\Rcm$), collective neutrino
  oscillations are suppressed by either the large matter density or
  the large  neutrino fluxes themselves. In this regime neutrinos can still
  experience $\nu$-enhanced MSW
  flavor transformation. Far away from the PNS ($r\gtrsim\Rcp$), where
  the neutrino fluxes are negligible, neutrinos experience either vacuum
  oscillations or conventional MSW flavor transformation, depending
  on the matter density and the energy of the neutrino. If $\Rcp>\Rcm$,
  then there exists a window ($\Rcm\lesssim r\lesssim\Rcp$, the
  ``bipolar regime'') where neutrinos experience collective 
  neutrino oscillations and where spectral swaps/splits develop. The
  curves show the  energy-averaged neutrino survival
  probabilities for electron neutrinos and antineutrinos in
  single-angle calculations chosen to be representative of some late-time
  supernova conditions.}
\end{figure*}

With the picture for neutrino oscillations developed above,
we can utilize the strength of neutrino self-interaction to
sketch out neutrino oscillation regimes.
These are shown in Figure~\ref{fig:Pnunu}.
We will focus the following discussion on the two flavor
mixing scenario with $|\dmsqr|\simeq\dmsqr_\mathrm{atm}$ and
$\thetav\simeq\theta_{13}$.   
We designate $\Rcm$ as
the radius closest to the PNS where collective oscillations set in.
Likewise, $\Rcp$ is the outer radius where collective oscillations cease.
In both the NH and IH cases,
$\Rcp$ can be estimated
from the condition 
\begin{equation}
\sqrt{2}\GF
n_{\bar\nu_e}(\Rcp)\simeq\frac{\dmsqr_\mathrm{atm}}{\langle
  E_{\bar\nu_e}\rangle},
\end{equation}
where $n_{\bar\nu_e}(r)$ is the total number density of the
antineutrinos at $r$ that are initially $\bar\nu_e$ at the neutrino
sphere, and $\langle E_{\bar\nu_e}\rangle$ is the average energy of
these antineutrinos. Here we have chosen to use $\bar\nu_e$ as the
representative neutrino species, and we have used
$\frac{\dmsqr_\mathrm{atm}}{\langle E_{\bar\nu_e}\rangle}$ as an estimate of
the frequency spread of the neutrino spectrum.

In the NH case, assuming a fully-synchronized neutrino system,
$\Rcm$ is approximately where \cite{Pastor:2002we}
\begin{equation}
\sqrt{2}\GF
n_e(\Rcm)\simeq\Omega_\sync. 
\end{equation}
(In the large
neutrino flux limit, the behavior of a neutrino system experiencing 
$\nu$-enhanced MSW flavor transformation is the same as that of a
synchronized neutrino system \cite{Duan:2007fw}.) 
We note that for an iron core collapse supernova at early times and in
the NH case,
$\Rcp\leq\Rcm$, meaning that the collective neutrino oscillation
regime does not exist.
At later times in these models
a collective neutrino oscillation regime can appear  
when the matter density becomes relatively small.
In contrast, collective neutrino oscillations can occur at early
epochs in an O-Ne-Mg core collapse supernova. This is because the
progenitors of these supernovae have relatively lower masses 
($8$--$12M_\odot$), and therefore, after core bounce, the PNS 
has a dilute, lower density envelope
\cite{Nomoto:1984aa,Nomoto:1987aa}.

In the IH case the value of $\Rcm$ is the larger of the values
extracted from the following two conditions:
\begin{equation}
n_{\bar\nu_e}(\Rcm) \simeq
\frac{1}{(\sqrt{1+\varepsilon}-1)^2} 
\frac{\Delta m_\mathrm{atm}^2}{\sqrt{2}\GF \langle E_{\bar\nu_e}\rangle}
\quad\text{and}\quad
n_{\bar\nu_e}(\Rcm) \simeq n_e(\Rcm).
\label{eq:Rcm}
\end{equation}
The first condition in Equation~\myeqref{eq:Rcm} is based on the flavor
pendulum model (See Equation~\myeqref{eq:mucr}) and the second condition
is in accord with the discussion in
Section~\ref{sec:decoherence}. 
In the very early epochs of an iron core collapse supernova,
the collective neutrino oscillation regime may not exist 
 in the IH case because of
the presence of a large matter density.

\subsection{Effects of collective neutrino oscillations}

The supernova neutrino energy spectra can be dramatically modified by
collective neutrino oscillations.
Perhaps the most prominent feature to arise from collective neutrino
oscillations is the spectral swap/split.
For example, see Figure~\ref{fig:P-theta-E} and its caption.
Obtaining this feature can depend on supernova conditions.
For example, the calculations used to produce Figure~\ref{fig:P-theta-E}
would give no spectral swap/split if the matter
density were too large \cite{Fogli:2007bk,EstebanPretel:2008ni}.

When $\nu_e$ and $\bar\nu_e$ are the most abundant
neutrino species at the neutrino sphere,
some qualitative aspects of collective neutrino
oscillations in supernovae can be understood using the flavor pendulum
model.
In the IH case, the initial configuration of the flavor
pendulum is near its highest position and is unstable when the
neutrino flux is below some critical value.
(See the discussion in Section~\ref{sec:swap}.) 
This implies that collective neutrino oscillations and
their effects on neutrino energy spectra are relatively insensitive to
the matter 
density or the exact value of $\theta_{13}$. In the NH
case, however, the flavor pendulum would be near its stable configuration unless
MSW flavor transformation displaces it from this position.
As a result, collective neutrino oscillations and their 
effects depend on the efficiency of the $\nu$-enhanced MSW flavor
transformation.
In turn, the efficiency of this flavor transformation
is sensitive to both the matter profile and the value of
$\theta_{13}$. This general picture is confirmed
by multi-angle calculations  \cite{Duan:2007bt}.
The results shown in Reference~\cite{Duan:2007bt} suggest that it may
be possible to resolve the neutrino mass hierarchy with an observed supernova
neutrino signal, even if the absolute neutrino masses and/or $\theta_{13}$ are
too small to be measured in the laboratory.

Neutrino signals detected at very early times after core bounce 
may be important probes of neutrino mixing.
For iron core collapse supernovae,
the neutrino spectra  at the very early times, 
are modified by the conventional MSW mechanism and can be easily calculated
\cite{Kachelriess:2004ds}.  For 
O-Ne-Mg core collapse supernovae, however, the matter density is so low
 above the PNS that even at these very early times,
where the $\nu_e$ luminosity can be very large ($\sim 10^{53}$ erg/s),
collective
neutrino oscillations can create step-like features (swaps/splits) in
the observed neutrino energy spectra \cite{Duan:2007sh}.
If these features are observed in a Galactic supernova neutrino burst, 
then they could serve as diagnostics of the neutrino mass
hierarchy and, in the NH case, even provide a measure of $\theta_{13}$.

However, we note that the spectral swap/split phenomenon is sensitive
to the neutrino luminosities and energy spectra at the neutrino
sphere. Presently there are rather large uncertainties in these quantities,
especially for the late-time supernova environment.
We also note
that very large neutrino fluxes do not
necessarily imply a large neutrino oscillation effect. In fact, in the
IH case a larger neutrino luminosity 
pushes $\Rcm$ to a larger radius.

Of course, collective neutrino
oscillations are not the only way that neutrinos may experience flavor
transformation in supernovae. As in the Sun,  in supernovae
neutrinos  also can experience
the conventional MSW flavor transformation. However, the matter 
profile in
supernovae  may not be smooth (e.g., because of shocks or turbulence)
near MSW resonance regions. This can produce some interesting
phenomena
\cite{Sawyer:1990tw,Loreti:1995ae,Schirato:2002tg,Fogli:2003dw,Kneller:2005hf,Friedland:2006ta,Dasgupta:2007aa,Duan:2009cd}.

In principle, the alteration of supernova neutrino energy spectra by 
collective neutrino oscillations and/or other processes of neutrino
flavor transformation  could affect supernova dynamics,
shock reheating, and nucleosynthesis in neutrino-heated ejecta
\cite{Fuller:1992aa,Qian:1993dg,Qian:1994wh,Sigl:1994hc,Chakraborty:2009ej}.
However, modeling of this physics in realistic supernova conditions
is at present primitive.


\section{Summary and open issues\label{sec:conclusions}}

In this review we have discussed collective neutrino oscillations
in a simple bipolar neutrino system, in homogeneous and isotropic
neutrino gases, and in anisotropic neutrino gases. The simple bipolar neutrino
system (described by the flavor pendulum) is the simplest of these
and is 
solvable analytically. It can be used to understand many qualitative features of
collective neutrino oscillations in supernovae. The single-angle
scheme essentially treats supernova neutrinos as a homogeneous and
isotropic gas, and adiabatic neutrino flavor transformation in
such a gas can be used to understand the spectral swap/split
phenomenon in the supernova environment. 
An anisotropic neutrino gas can
possess unique characteristics (e.g., suppression of collective
flavor oscillations by large matter density).
This makes it an important target for study
because realistic physical environments such as core collapse
supernovae can be highly anisotropic.

We have covered some of the basic properties of
collective neutrino oscillations, with emphasis on the two-flavor mixing case. 
There is much about the collective neutrino
oscillation phenomenon which remains to be understood.
For example, in Section~\ref{sec:twosols}
we skipped over the adiabaticity condition when we discussed adiabatic
neutrino flavor transformation. 
Adiabaticity criteria developed so far \cite{Raffelt:2007xt}
are difficult to use in practice.

\begin{figure*}
\begin{center}        
  \includegraphics*[width=0.6 \textwidth, keepaspectratio]{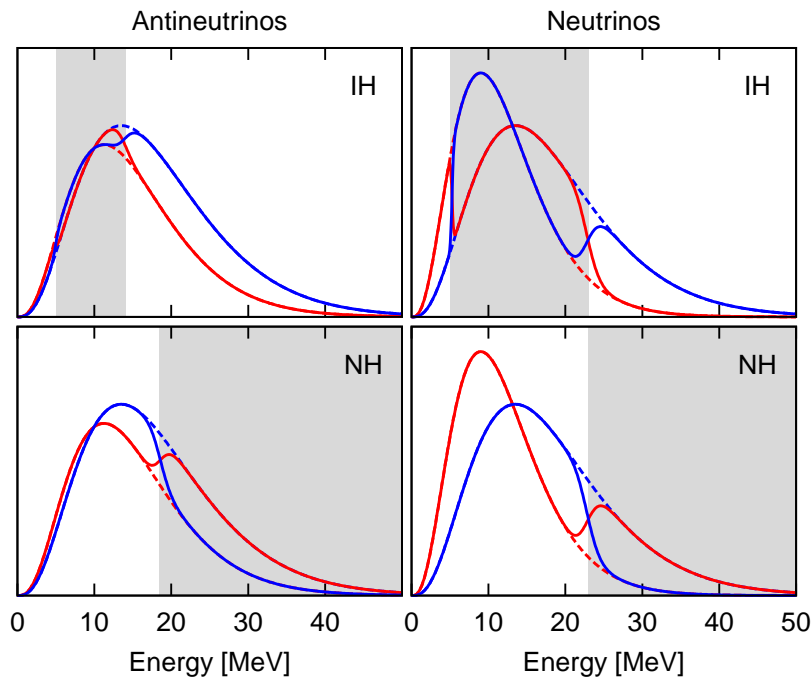}
\end{center}
\caption{\label{fig:multisplits}Multiple spectral swaps/splits in a
  two-flavor single-angle calculation with
  $|\dmsqr|\simeq\dmsqr_\mathrm{atm}$ and $\thetav\ll1$. The solid and
  dashed curves are 
  the initial and final energy spectra, respectively. The red and blue curves
  are for $e$ and $\mu$ flavors, respectively. The shaded region mark
  the energy ranges where spectral swaps occur. 
  Unlike previous numerical calculations, this calculation assumes
  that $\nu_\mu$ 
  and $\bar\nu_\mu$, rather than $\nu_e$ and $\bar\nu_e$, are the most
  abundant neutrino species at the neutrino sphere.
  Reprinted figure with permission from 
  \href{http://prl.aps.org/abstract/PRL/v103/i5/e051105}{B.~Dasgupta et al, 
  Phys.\ Rev.\ Lett., Vol.\ 103, 051105, 2009} 
  (Reference~\cite{Dasgupta:2009mg}). 
  Copyright 2009 by the American Physical Society.} 
\end{figure*}

 There are other open issues in  our current
understanding of collective neutrino oscillations in supernovae.
For example, it was recently shown that there can
exist multiple spectral 
swaps/splits in the final neutrino energy spectra
\cite{Fogli:2007bk,Dasgupta:2009mg,Fogli:2009rd}
(Figure~\ref{fig:multisplits}).
The appearance of these features depend on, however, the
luminosities and energy spectra of the different neutrino species at
the neutrino sphere. This finding
cannot be explained by a grand collective precession mode
in which all neutrinos and antineutrinos participate.
In an even more recent paper \cite{Friedland:2010sc} it was
reported that qualitatively different results could appear in some
collective neutrino oscillation scenarios depending on whether 
two-flavor mixing or full three-flavor mixing is used. All of these
discoveries point up the need for a systematic study of 
supernova neutrino oscillations with neutrino self-interaction
and full three-flavor mixing at various supernova epochs
where neutrino luminosities and energy spectra can be 
different. In addition, numerical simulations of supernova explosions have
 shown clearly that realistic supernova environments may be highly
anisotropic and inhomogeneous
\cite{Herant:1994dd,Fryer:2002zw,Buras:2005rp,Blondin:2005wz,Blondin:2006yw,Bruenn:2007bd,Ott:2008jb,Hammer:2009cn}. 
It would be interesting to see how 
density inhomogeneities arising from, e.g., shock waves and turbulence
might affect 
collective neutrino oscillations. It remains a towering numerical challenge to
integrate fully hydrodynamic, three-dimensional supernova models with 
calculations of neutrino oscillations that include neutrino self-interaction. 
However, such self-consistent integration may be necessary in order to
study the interplay between supernova
physics and neutrino oscillations.
In particular, a full
understanding of neutrino oscillations in supernovae may hold the key to 
deciphering the neutrino signal from a future Galactic supernova.
The stakes are high. 
Deciphering a supernova neutrino signal 
could provide important insights into supernova astrophysics.
It also could provide key insights
into fundamental neutrino properties and these insights could be
complementary to those sought by future neutrino experiments.

\begin{acknowledgments}
The authors would like to thank J.~Carlson, J.~Cherry, A.~Friedland, W.~Haxton,
D.~Kaplan, C.~Kishimoto, A.~Kusenko, A.~Mezzecappa, G.~Raffelt and
S.~Reddy for valuable 
conversations. This work was 
supported in part by the Institutional Computing Program at LANL and
the National Energy Research Scientiﬁc Computing Center, which is
supported by the DOE Office of Science under contract
no.~DE-AC02-05CH11231. This work was also 
supported by DOE grants DE-FG02-00ER41132 at the INT,
DE-FG02-87ER40328 at UMN, NSF grant PHY-06-53626 at UCSD and an
IGPP/LANL mini-grant. The research of H.~D.\  is supported by LANL
LDRD program through the Director's postdoctoral fellowship at LANL.
\end{acknowledgments}

\bibliography{ref}

\end{document}